\documentclass[usenatbib]{mnras}
\usepackage[T1]{fontenc}
\usepackage[latin9]{inputenc}
\usepackage{array}
\usepackage{amsmath}
\usepackage{graphicx}
\usepackage{esint}
\usepackage[authoryear]{natbib}

\makeatletter

\providecommand{\tabularnewline}{\\}

\makeatother

\begin{document}
\title{On effects of surface bipolar magnetic regions on the convection zone
dynamo}
\author[V.V.~Pipin]{V.V.~Pipin,\thanks{email: pip@iszf.irk.ru}\\
 Institute of Solar-Terrestrial Physics, Russian Academy of Sciences,
Irkutsk, 664033, Russia }
\maketitle
\begin{abstract}
We investigate the effect of the surface bipolar magnetic regions
(BMR) on the large-scale dynamo distributed in the bulk of the convection
zone. The study employs the nonlinear 3D mean-field dynamo model.
We model the emergence of the BMR on the surface through the nonaxisymmetric
magnetic buoyancy effect, which acts on the large-scale toroidal magnetic
field in the upper half of the convection zone. The nonaxisymmetric
magnetic field which results from this mechanism is shallow. On the
surface, the effect of the BMR on the magnetic field generation is
dominant. {However, because of the shallow BMR distribution, its effect
on the global dynamo is less compared to the convective zone
dynamo.} We find that the mean-field $\alpha$
effect, which acts on the nonaxisymmetric magnetic field of the BMRs,
provides the greater contribution to the dynamo process than the BMR's
tilt does. Even so, the fluctuations of the BMR's tilt lead to the
parity braking in the global dynamo. At the surface the nonaxisymmetric
magnetic field, which are generated because of the BMR's activity,
shows a tendency for the bihelical spectrum with the positive sign
for the low $\ell$ modes during the maximum of the magnetic activity
cycle.
\end{abstract}

\section{Introduction}

Parker (1955) represented the whole dynamo process as the dynamo waves
propagating through the convection zone. This idea grew to the mean-field
MHD and dynamo theory (\citet{Moffatt1978,Krause1980}). The mean-field
theory framework employs the idea of the scale separation between
the mean and turbulent parts of the magnetic field and flow. It assumes
that the mean large-scale field can be extracted from the background
turbulent astrophysical plasma using an appropriate averaging procedure
(see the above-cited textbooks). In the solar-type dynamos, the axisymmetric
magnetic field dominates. On the Sun, the large-scale nonaxisymmetric
dynamo is well below the dynamo instability threshold because of differential
rotation (\citet{Raedler1986}). Yet, most of the energy of the surface
magnetic field is concentrated on small scales \citep{Vidotto2018}.
The origin of the surface nonaxisymmetric magnetic fields seems to
be connected with the emergence and decay of solar active regions
\citep{Wang1989b}.

The effect of the surface magnetic activity on the large-scale dynamo
is poorly understood. One way to account for it is through the boundary
conditions. Under the insulator boundary condition, the external magnetic
field is potential, and the toroidal magnetic field is zero at the
surface. Observations show that more realistic boundary conditions
should allow penetration of the toroidal magnetic field to the surface
(\citet{Moss1992,Bonanno2016}). Note that the penetration of the
toroidal magnetic field to the surface allows the efficient poloidal
magnetic field generation in the near-surface layer.

The Babcock-Leighton scenario accounts for the surface magnetic activity
on the dynamo phenomenologically (\citealp{Babcock1961} and
\citealp{Leighton1964}). The phenomenological picture behind it was
established and summarized in several papers and reviews (see, e.g.,
\citealp{Giovanelli1985,Wang1989b,Petrie2014}). This scenario is
employed in the flux transport dynamo models ( see \citealp{Mackay2012,Ugarte-Urra2015,Cameron17}
and references therein ). In this scenario, the tilted BMRs emerging
and developing at the surface result in the dynamo generation of the
poloidal magnetic field of the Sun. In the convection zone dynamo
scenario, the regeneration of the poloidal magnetic field goes everywhere
in the depth of the convection zone.

Results of the flux transport dynamo models (see, \citealt{Karak2014a,Passos2014,Hazra2019n})
show the necessity of the mean-field effects to account for the possible
complicated properties of the global flow and properties of the long-term
variations of the solar activity. Yet, the simple framework of the
Babcock-Leighton scenario faces problems to reproduce the results
of the global convection dynamo simulation (see, \citet{Schrinner2011,Schrinner2011a,Warnecke2021}).
This is because of a poor representation of the mean electromotive
force of turbulent flows and magnetic fields. In this paper, we consider
the mean-field dynamo model of \citet{Pipin2019c}. To account for
the dynamo effects of bipolar magnetic regions, the model prescribes
the nonaxisymmetric magnetic buoyancy and $\alpha$ effects acting
on the buoyant part of the toroidal magnetic field.

\section{Model}

In this study, we employ the mean-field MHD framework, which was formulated
by \citet{Roberts1975} and \citet{Krause1980} to describe the evolution
of the large-scale magnetic field in the turbulent astrophysical plasma.
The model solves the dynamo equations together with the equations
governing the angular momentum balance, meridional circulation and
the mean heat transport in the convection zone.

\subsection{Mean-field dynamo equations}

The magnetic field evolution is governed by the mean-field induction
equation: 
\begin{equation}
\partial_{t}\left\langle \mathbf{B}\right\rangle =\mathbf{\nabla}\times\left(\mathbf{\mathbf{\boldsymbol{\mathbf{\mathcal{E}}}}+}\left\langle \mathbf{U}\right\rangle \times\left\langle \mathbf{B}\right\rangle \right)\,,\label{eq:mfe}
\end{equation}
where $\mathbf{\mathcal{E}}=\left\langle \mathbf{u\times b}\right\rangle $
is the mean electromotive force; $\mathbf{u}$ and $\mathbf{b}$ are
the turbulent fluctuating velocity and magnetic field, respectively;
and $\left\langle \mathbf{U}\right\rangle $ and $\left\langle \mathbf{B}\right\rangle $
are the mean velocity and magnetic field. We assume that the averaging
is done over the ensemble of turbulent flows and magnetic fields.
Like \citet{Moss1991}, we represent the vector $\left\langle \mathbf{B}\right\rangle $
by sum of the axisymmetric and nonaxisymmetric parts. These parts
are further decomposed into a sum of the poloidal and toroidal components,
as follows: 
\begin{eqnarray}
\left\langle \mathbf{B}\right\rangle  & = & \overline{\mathbf{B}}+\tilde{\mathbf{B}}\,,\label{eq:b0}\\
\mathbf{\overline{B}} & = & \hat{\mathbf{\phi}}B+\nabla\times\left(A\hat{\mathbf{\phi}}\right)\,,\label{eq:b1}\\
\tilde{\mathbf{B}} & = & \mathbf{\nabla}\times\left(\mathbf{r}T\right)+\mathbf{\nabla}\times\mathbf{\nabla}\times\left(\mathbf{r}S\right),\label{eq:b2}
\end{eqnarray}
where $\overline{\mathbf{B}}$ and $\tilde{\mathbf{B}}$ are the axisymmetric
and nonaxisymmetric components of the large-scale magnetic field.
Also, $\hat{\mathbf{\phi}}$ is the azimuthal unit vector, $\mathbf{r}$
is the radius vector, $r$ is the radial distance, and $\theta$ is
the polar angle. The gauge transformation for superpotentials $T$ and $S$ involves a sum with the arbitrary r-dependent function (\citealt{Krause1980}).
We seek the solution for ${T}$ and ${S}$ using the spherical harmonic
decomposition. In this case, by definition, the zero $\ell$ modes
are excluded. In this case, the representation of the Eq(\ref{eq:b2})
is gauge invariant (also, see, \citealt{Berger2018}). We assume that
the large-scale flow is axisymmetric $\left\langle \mathbf{U}\right\rangle \equiv\overline{\mathbf{U}}$.
Similarly, we assume that the mean entropy and the other thermodynamic parameters are axisymmetric, as well. Nevertheless, the model takes into account the effect of
the nonaxisymmetric magnetic field on the angular momentum balance
and the meridional circulation using the longitudinal averaging of
the Lorentz force. To get the dynamo equations for the axisymmetric
toroidal magnetic field evolution, we take the scalar product of the Eq(\ref{eq:mfe})
with the unit vector $\hat{\phi}$. We do the same for the uncurled version
the Eq(\ref{eq:mfe}) to get the equation for the vector potential
$A$. To get the evolution equations for the nonaxisymmetric magnetic field we  take  the curl and double curl of the Eq(\ref{eq:mfe}).
Then, we take the scalar product of these equations with vector $\mathbf{r}$
(see, details in \citealt{Krause1980}, and \citealt{Moss1991}).

We decompose the mean electromotive force for two parts: 
\begin{eqnarray}
\mathcal{E}_{i} & = & \mathcal{E}_{i}^{(A)}+\mathcal{E}_{i}^{(\mathrm{BMR})}\,,\label{eq:EMF-1}
\end{eqnarray}
where the expression for $\mathcal{E}_{i}^{(A)}$ results from the analytical computations (see, e.g., \citealt{Kitchatinov1994,Pipin2008a}).
The $\mathcal{E}_{i}^{(\mathrm{BMR})}$ stands for the phenomenological
part of mean electromotive force. I introduce it to take the effects
of the BMRs into account. 

The $\mathcal{E}_{i}^{(A)}$ reads as follows, 
\begin{equation}
\mathcal{E}_{i}^{(A)}=\left(\alpha_{ij}+\gamma_{ij}\right)\left\langle B\right\rangle _{j}-\eta_{ijk}\nabla_{j}\left\langle B\right\rangle _{k},\label{eq:Ea}
\end{equation}
where the tensor, $\alpha_{ij}$ stands for the turbulent generation
by $\alpha$-effect, $\gamma_{ij}$ is the turbulent pumping and $\eta_{ijk}$
is the eddy magnetic diffusivity tensor. The analytical expressions
of the $\alpha_{ij}$, $\gamma_{ij}$ and $\eta_{ijk}$ take into account
effects of the global rotation, magnetic field and density stratification
on the turbulent convection. The same model was recently considered
by \citet{Pipin2019c} and \citet{Pipin2020}. For convenience, we
present the full expressions of the above tensors in Appendix A.

The $\alpha$ effect tensor includes the small-scale magnetic helicity
density contribution, i.e., the pseudo scalar $\left\langle \chi\right\rangle =\left\langle \mathbf{a}\cdot\mathbf{b}\right\rangle$ (where
$\mathbf{a}$ and $\mathbf{b}$ are the fluctuating vector-potential
and magnetic field, respectively), 
\begin{eqnarray}
\alpha_{ij} & = & C_{\alpha}\psi_{\alpha}(\beta)\alpha_{ij}^{(H)}+\alpha_{ij}^{(M)}\psi_{\alpha}(\beta)\frac{\left\langle \chi\right\rangle \tau_{c}}{4\pi\overline{\rho}\ell_{c}^{2}},\label{alp2d}
\end{eqnarray}
where, the contributions of the kinetic $\alpha$ effect tensor $\alpha_{ij}^{(H)}$
and the magnetic helicity effect tensor $\alpha_{ij}^{(M)}$ are given
in Appendix A. The radial profiles of the $\alpha_{ij}^{(H)}$ and
$\alpha_{ij}^{(M)}$ depend on the mean density stratification, the profile
of the convective RMS velocity $u_{c}$ and on the Coriolis number
$\Omega^{*}=2\Omega_{0}\tau_{c}$, where $\Omega_{0}$ is the angular
velocity of the star and $\tau_{c}$ is the convective turnover time.
The magnetic quenching function $\psi_{\alpha}(\beta)$ depends on
the parameter $\mathrm{\beta=\left|\left\langle \mathbf{B}\right\rangle \right|/\sqrt{4\pi\overline{\rho}u_{c}^{2}}}$
(see Appendix A). Note that in the presence of the $\tilde{\mathbf{B}}$-field,
the $\alpha$ effect tensor becomes nonaxisymmetric. This effect is caused both by the $\psi_{\alpha}(\beta)$-quenching and the magnetic helicity effects.

The magnetic helicity evolution follows the global conservation law
for the total magnetic helicity, $\left\langle \chi\right\rangle ^{(tot)}=\left\langle \chi\right\rangle +\left\langle \mathbf{A}\right\rangle \cdot\left\langle \mathbf{B}\right\rangle $,
(see, \citet{Hubbard2012,Pipin2013c,Brandenburg2018}): 
\begin{equation}
\left(\frac{\partial}{\partial t}+\boldsymbol{\left\langle \mathbf{U}\right\rangle \cdot\nabla}\right)\left\langle \chi\right\rangle ^{(tot)}=-\frac{\left\langle \chi\right\rangle }{R_{m}\tau_{c}}-2\eta\left\langle \mathbf{B}\right\rangle \cdot\left\langle \mathbf{J}\right\rangle -\mathbf{\nabla\cdot}\mathbf{\mathbf{\mathcal{F}}}^{\chi},\label{eq:helcon}
\end{equation}
where, we use ${\displaystyle 2\eta\mathbf{\left\langle b\cdot j\right\rangle }=\frac{\left\langle \chi\right\rangle }{R_{m}\tau_{c}}}$
\citep{Kleeorin1999}. Also, we introduce the diffusive flux of the
small-scale magnetic helicity density, $\mathbf{\mathbf{\mathcal{F}}}^{\chi}=-\eta_{\chi}\mathbf{\nabla}\left\langle \chi\right\rangle $,
and $R_{m}$ is the magnetic Reynolds number. The coefficient of the
turbulent helicity diffusivity, $\eta_{\chi}$, is chosen ten times
smaller than the isotropic part of the magnetic diffusivity \citet{Mitra2010}:
$\eta_{\chi}=\frac{1}{10}\eta_{T}$.

The mean magnetic helicity density is formally decomposed into the
axisymmetric and nonaxisymmetric parts: $\left\langle \chi\right\rangle ^{(tot)}=\overline{\chi}^{(tot)}+\tilde{\chi}^{(tot)}$.
The same is done for the magnetic helicity density of the turbulent
field: $\left\langle \chi\right\rangle =\overline{\chi}+\tilde{\chi}$,
here $\overline{\chi}=\overline{\mathbf{a}\cdot\mathbf{b}}$ and $\tilde{\chi}=\tilde{\left\langle \mathbf{a}\cdot\mathbf{b}\right\rangle }$.
Thus, we have, 
\begin{eqnarray}
\overline{\chi}^{(tot)} & = & \overline{\chi}+\overline{\mathbf{A}}\cdot\overline{\mathbf{B}}+\overline{\tilde{\mathbf{A}}\cdot\tilde{\mathbf{B}}},\label{eq:t1}\\
\tilde{\chi}^{(tot)} & = & \tilde{\chi}+\overline{\mathbf{A}}\cdot\tilde{\mathbf{B}}+\tilde{\mathbf{A}}\cdot\overline{\mathbf{B}}+\tilde{\mathbf{A}}\cdot\tilde{\mathbf{B}},\label{eq:t2}
\end{eqnarray}
The evolution of the $\overline{\chi}$ and $\tilde{\chi}$ is governed
by the corresponding parts of Eq(\ref{eq:helcon}). The magnetic helicity
conservation is determined by the magnetic Reynolds number $R_{m}$.
In this paper, we employ $R_{m}=10^{6}$.

\subsection{The BMR's formation and its dynamo effects}

To take into account the effects of the surface bipolar magnetic regions
(BMRs) on the dynamo, we introduce the phenomenological part of the
mean electromotive force as follows, 
\begin{equation}
\mathcal{E}_{i}^{(\mathrm{BMR})}=\alpha_{\beta}\delta_{i\phi}\left\langle B\right\rangle _{\phi}+V_{\beta}\left(\hat{\boldsymbol{r}}\times\left\langle \mathbf{B}\right\rangle \right)_{i},\label{eq:ep}
\end{equation}
where the first term takes into account the BMR's $\alpha$-effect
and the second term does the same for the magnetic buoyancy effect.
Our motivation for the Eq.(\ref{eq:ep}) following \citet{Parker1979}
idea. We assume that some part of the toroidal magnetic field in the
upper part of the convection zone becomes buoyantly unstable, it emerges
and forms the surface BMR. {The second term of the
  Eq.(\ref{eq:ep}) describes the buoyant emergence of the BMR. Later we
will see, that the first term of this formula is related with the
BMR's tilt.} A mechanism of the instability can be rather complicated
(see, e.g., \citealt{Gilman1970,Gilman2018}). Also, it is likely,
that other processes in the solar convection zone can be responsible
for the BMRs formations (see, e.g.,\citealt{Kleeorin1989,Mazur2000SoPh,Getling2001,Brandenburg2013,Stein2012,Losada2017,Kleeorin2020}).
Here, the model employs the magnetic buoyancy effect formally. Its
main goal is to mimic the BMRs on the solar surface. Other implementations
of the BMRs can be found in the models of the Babcock-Leighton type
(see, e.g., \citealt{2008ApJ673544Y,Brun2014,Miesch2014Ap}).

The magnetic buoyancy velocity is modeled using the turbulent and mean-field buoyancy effects. Its expression was suggested ealier (see,  \citealp{Kitchatinov1992,Kitchatinov1993,Ruediger1995}). Following these results we put, 
\begin{eqnarray}
V_{\beta} & = & V_{m}\xi_{\beta}(t,\boldsymbol{r})\label{eq:bu}\\
V_{m} & = & \frac{\alpha_{\mathrm{M}}u_{c}}{\gamma}\mathcal{H}\left(\beta_{m}\right),
\end{eqnarray}
where $\mathrm{\alpha_{M}}=1.9$ is the mixing-length theory parameter,
$\gamma$ is the adiabatic law constant, $u_{c}$ is the convective
RMS velocity and $\mathrm{\beta=\left|\left\langle \mathbf{B}\right\rangle \right|/\sqrt{4\pi\overline{\rho}u_{c}^{2}}}$.
The function $\mathcal{H}\left(\beta\right)$ takes into account the
effect of the magnetic tensions on the mean-field magnetic buoyancy
(see, Appendix A). The subscript 'm' in the Eq(\ref{eq:bu}) marks
that the amplitude of the effect is taken at the location of maximum
the magnetic field strength. Note, that the turbulent pumping tensor,
$\gamma_{ij}$, in the Eq.(\ref{eq:Ea}) take into account the mean-field
magnetic buoyancy, as well, (though there we have $\xi_{\beta}=0$
and smooth profile of $\mathcal{H}\left(\beta\right)$ corresponding
to the B-field distribution). The spatial and temporal parameters
of the $\xi_{\beta}$ are controlled by the formula, 
\begin{equation}
\!\xi_{\beta}\left(\boldsymbol{r},t\right)\!=\!\!\psi(r,t)\negthinspace\exp\left(\!\!-m_{\beta}\left(\!\sin^{2}\!\left(\!\frac{\phi\!-\!\phi_{m}}{2}\!\right)\!\!+\!\!\sin^{2}\!\left(\!\frac{\theta\!-\!\theta_{\mathrm{m}}}{2}\!\right)\!\right)\!\right)\!,\label{xib}
\end{equation}
where $\psi$ is a kink type function of radius and time, 
\begin{eqnarray}
\psi\! & =\!\! & \frac{1}{2}\left(\!1\!-\!\mathrm{erf}\left(50\left(r-r_{m}\right)\right)\!\right)\!\mathrm{e}\!^{{\displaystyle \frac{t}{\tau_{0}}}}\!,t\!<\!\delta t\label{kink}\\
 & = & 0,t>\delta t\,.,\nonumber 
\end{eqnarray}
where $r_{m}$ and $\theta_{m}$ are the radius and the latitude of
the toroidal magnetic field strength extrema in the upper part of
the convection zone. {The other parameters in the Eqs(\ref{xib},\ref{kink})
are as follows. We put the emergence time, $\delta t$ to 5 days.
The parameter $\tau_{0}$ controls the growth rate of the BMRs. In
our simulations, we put $\tau_{0}$=1 day, which roughly corresponds
to the results of \citet{Stenflo2012a}). The longitudinal
coordinate $\phi_{m}$ is random. The similar parameters of $\delta t$
and $\tau_{0}$ are employed in the spotmaker module of the 3D Babckock-Leighton
model of \citet{Miesch2014Ap}. The size of the BMR is controlled
by the parameter, $m_{\beta}$. In the paper, we put $m_{\beta}=100$.
This results in about 10$^{\circ}$ separation between the leading
and following polarity of the BMR. The nonaxisymmetric perturbations,
$\xi_{\beta}$, are randomly initiated in time and longitude in each
hemisphere independently. The parameters $\delta t$, $\tau_{0}$,
and $m_{\beta}$ affect the magnetic flux of the BMR and the total
magnetic flux generated by the BMRs. In our model, the magnitude of
the magnetic flux of the typical BMR is around 4-5$\cdot10^{22}$
Mx. The decrease of $\tau_{0}$, and $m_{\beta}$ and the increase
of $\delta t$ results in the increase of the magnetic flux of the
emerging BMR. }

In following of \citet{Parker1979}, it is assumed that the large-scale
toroidal magnetic field becomes unstable, when its strength decreases
outward faster than the mean density does. In particular, we compute
the parameter 
\begin{equation}
I_{\beta}=-r\frac{\partial}{\partial r}\log\frac{\left|\overline{B}\right|^{\zeta}}{\overline{\rho}},\label{eq:inst}
\end{equation}
where $\overline{B}$ is the strength of the axisymmetric toroidal
magnetic field and $\overline{\rho}$ is the density profile. For
the case of $\zeta=1$ we get the Parker's instability condition.
In this case, we find that the BMR's productivity is not enough to
reproduce the solar observations. In our simulations we use $\zeta=1.2$.
The dependence of $\zeta$ parameter on the physical conditions in
the stellar convection zone deserves a separate study and it is out
of our scope for this paper. The $\xi_{\beta}\left(\boldsymbol{r},t\right)$
is initiated when and if $I_{\beta}\left(r_{m},\theta_{m}\right)>0$
. In addition, we restrict the instability region using the following
conditions: $r_{m}>0.85R$ and $\left|\overline{B}_{m}\right|>500$G. 

\begin{figure}
\includegraphics[width=0.9\columnwidth]{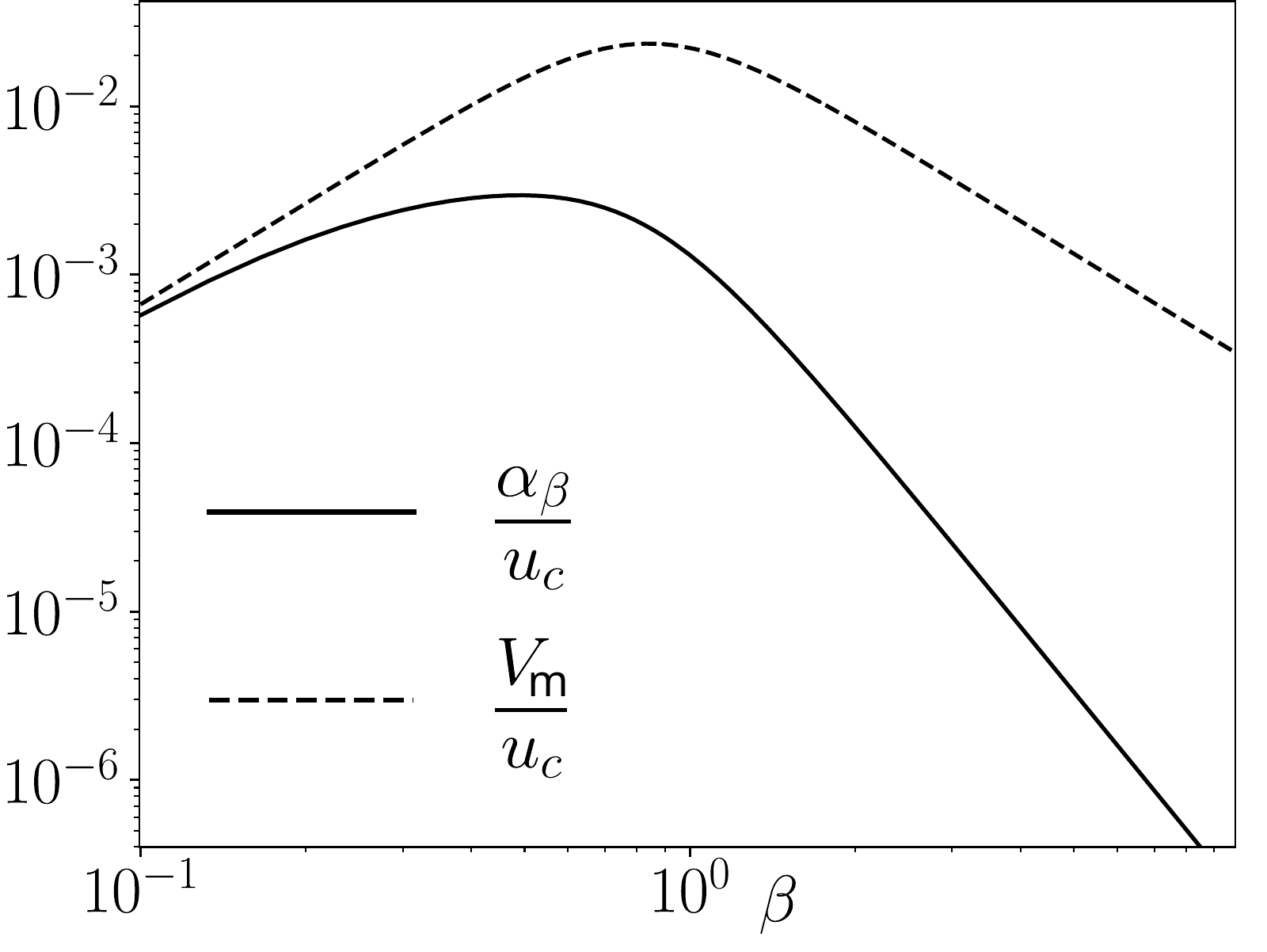}

\caption{\label{fig:cb}Connection of the mean-field buoyancy velocity (dashed
line) and the $\alpha_{\beta}$ effect (solid line) with the magnetic
field strength parameter $\beta$.}
\end{figure}

The $\alpha$-effect of the $\boldsymbol{\mathcal{E}}^{(\mathrm{BMR})}$
is given as follows 
\begin{equation}
\alpha_{\beta}=C_{\alpha\beta}\left(1+\xi_{\alpha}\right)\cos\theta V_{\beta}\psi_{\alpha}(\beta).\label{eq:ab}
\end{equation}
Here, we put the amplitude of the $\alpha$-effect to be determined
by the local magnetic buoyancy velocity. The $\xi_{\alpha}$ parameter
controls the random fluctuation of the BMR's $\alpha$-effect. The
parameter, $C_{\alpha\beta}$ controls the amplitude of the BMR's
$\alpha$-effect and tilt in the different simulation runs (see the
Subsection 2.4). The Fig\ref{fig:cb} shows that the nonlinear profiles
of the $\mathcal{H}\left(\beta\right)$ and $\psi\left(\beta\right)$
put the limits on the magnitude of the possible buoyancy velocity
drift and the corresponded $\alpha_{\beta}$ effect. The velocity
drift is order of $V_{m}\sim$$\frac{1}{40}u_{c}$ for the $\beta\sim1$
and the $\alpha_{\beta}$ is the order of magnitude smaller. The mean-field
solar dynamo models of \citet{Pipin2020} operates in the weakly nonlinear
regime where $\beta\le0.2$. In this case, $V_{m}\sim10^{-3}u_{c}$.

\begin{figure}
\includegraphics[width=0.95\columnwidth]{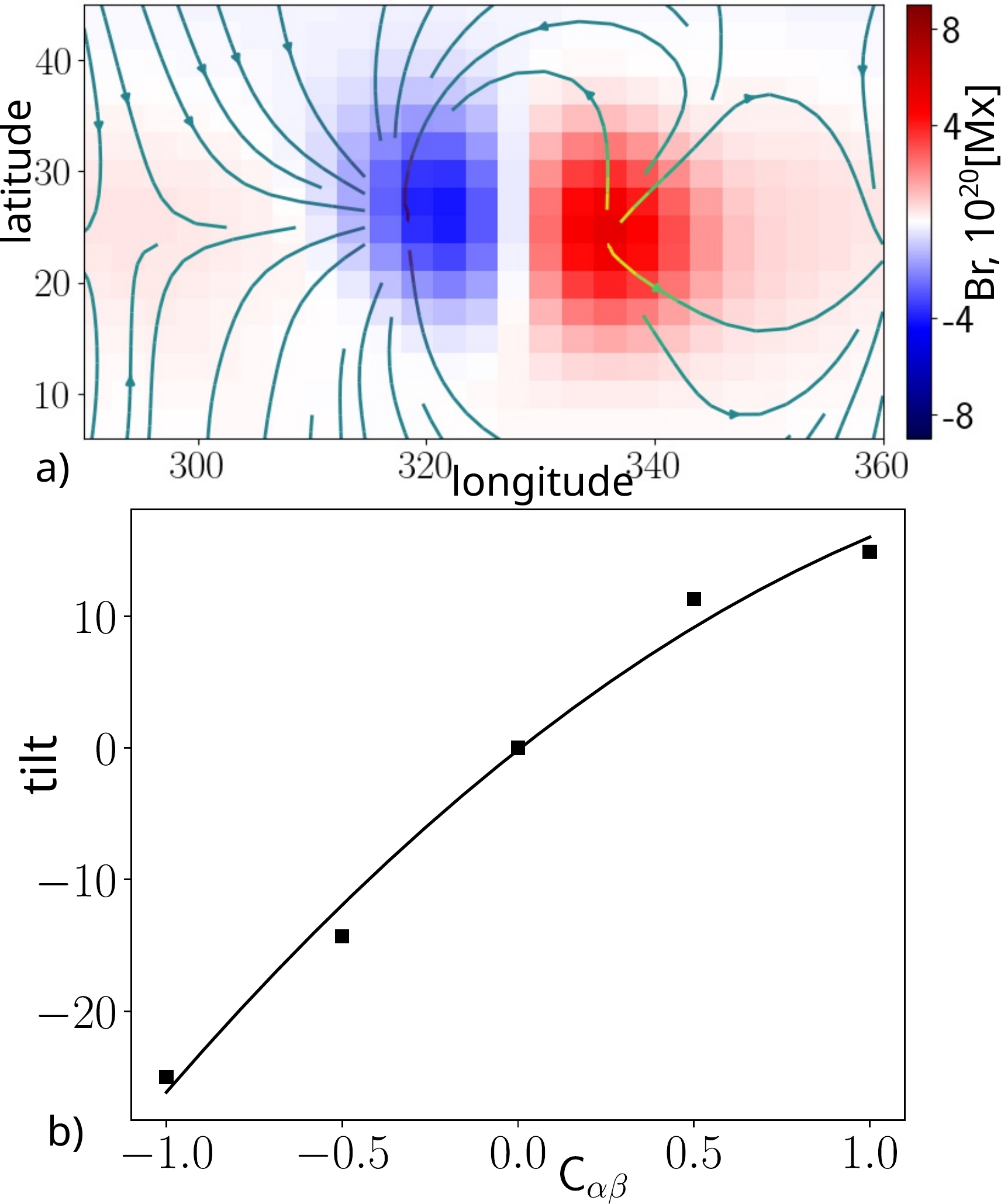}

\caption{\label{fig:tilt}a) Snapshot of the bipolar group at the surface for
$C_{\alpha}=1$, the color image shows the flux of the radial magnetic
field and streamlines show the surface components of the magnetic
field; b) Relation of the BMRs tilt and the parameter $C_{\alpha\beta}$
for the latitude 25$^{\circ}$.}
\end{figure}

\begin{figure}
\includegraphics[width=0.95\columnwidth]{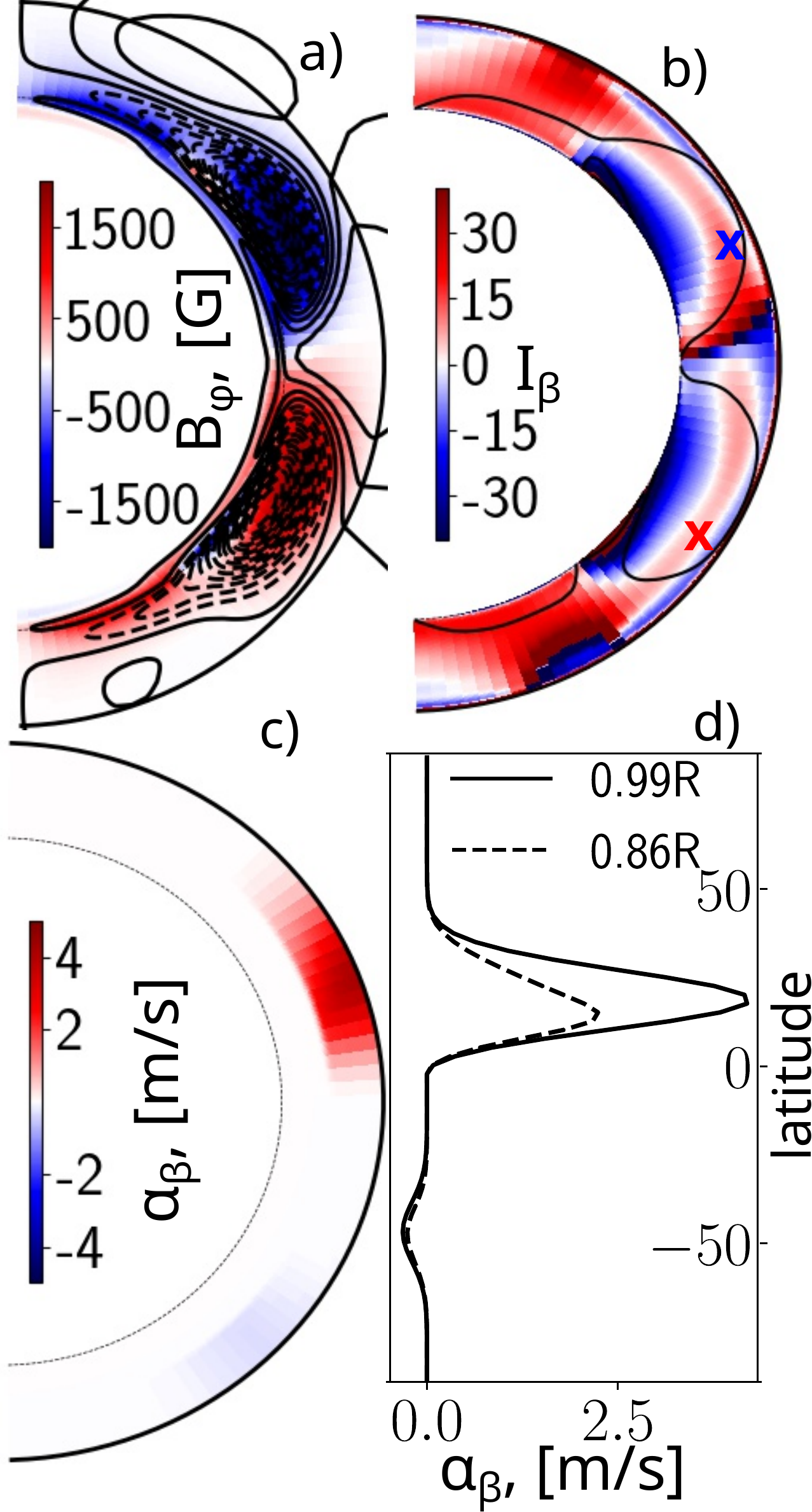}

\caption{\label{fig:sm}a) Snapshot of the axisymmetric magnetic field distribution
for the growing phase of the magnetic cycle. The color shows the toroidal
magnetic field strength, streamlines show the poloidal magnetic field;
b) the snapshot of the instability parameter, $I_{\beta}$ ; the red
and blue crosses show {locations of the unstable regions ($I_{\beta}\left(r_{m},\theta_{m}\right)>0$,
$r_{m}>0.85R$ and $\left|\overline{B}_{m}\right|>500$G) }in the
southern and northern hemispheres, respectively; c) the snapshot of
the $\alpha$ effect parameter, $\alpha_{\beta}$, for $C_{\alpha\beta}=0.5$
and $\xi_{\beta}\left(r,\delta t=5\tau_{0}\right)$; d) the latitudinal
profiles of the $\alpha_{\beta}$ at the bottom of the instability
region (dashed line) and at the top (solid line).}
\end{figure}

The Fig.\ref{fig:tilt} shows the relation between the magnitude of
the BMR's $\alpha$-effect and tilt for the latitude 25$^{\circ}$.
To calculate this relation, we used the short runs. Each run starts with 
the pure axisymmetric magnetic field distribution, which is shown
in the Fig\ref{fig:sm}a. The BMR is injected in the northern hemisphere
of the Sun. We assume that the axis of the BMR connects the extrema
points of opposite polarities. The positive tilt corresponds to Joy's
law. For the case $C_{\alpha\beta}=0.5$, the tilt of the BMRs is
about 10$^{\circ}$ at 25$^{\circ}$ latitude. This agrees with the
results of \citet{Tlatov2013}. Therefore, we choose $C_{\alpha\beta}=0.5$
for the long-term runs of the model. Figure \ref{fig:tilt}b shows
the relation between the amplitude of the $\alpha_{\beta}$ and the
BMR's tilt. The magnitude of tilt for the positive $C_{\alpha\beta}$
is less than for the negative ones. This is likely because of the
magnetic helicity effect which is contributed by the third term of
the Eq(\ref{eq:maina}). A detailed investigation of this effect is out of our scope.

We notice that the $\alpha$ effect of the BMRs is readily linked
with the tilt. The latitudinal dependence of this relationship is
governed by the factor $\cos\theta$, see the Eq.(\ref{eq:ab}). 

Figure \ref{fig:sm} shows snapshots of the axisymmetric magnetic,
as well as distribution of the instability parameter $I_{\beta}$
and results of calculation of the $\alpha_{\beta}$ parameter for
the distributions of the large-scale magnetic field. We take
these snapshots at the growing phase of the magnetic cycle. In the
southern hemisphere, the position of the dynamo wave is closer to
the equator and the surface than in the northern hemisphere. {The
structure of the dynamo wave and the shallow instability effect ($r_{m}>0.85$R,
see comments below the Eq(\ref{kink})) result in the difference in
positions of the unstable points of the dynamo waves in the South
and North hemispheres. The amplitude of the $\alpha_{\beta}$ parameter
is different as well.} In the northern hemisphere, the magnitude of
the $\alpha_{\beta}$ is close to the maximum magnitude of the kinetic
$\alpha$ effect (cf, Fig\ref{fig1}c). 

We model the randomness of the tilt using the parameter $\xi_{\alpha}$.
Similar to \citet{Rempel2005c}, the $\xi_{\alpha}$ evolution follows
the Ornstein--Uhlenbeck process, 
\begin{eqnarray}
\dot{\xi}_{\alpha} & = & -\frac{2}{\tau_{\xi}}\left(\xi_{\alpha}-\xi_{1}\right),\label{xia}\\
\dot{\xi}_{1} & = & -\frac{2}{\tau_{\xi}}\left(\xi_{1}-\xi_{2}\right),\nonumber \\
\dot{\xi}_{2} & = & -\frac{2}{\tau_{\xi}}\left(\xi_{2}-g\sqrt{\frac{2\tau_{\xi}}{\tau_{h}}}\right).\nonumber 
\end{eqnarray}
Here, $g$ is a Gaussian random number. It is renewed every time step,
$\tau_{h}$ . The $\tau_{\xi}$ is the relaxation time of $\xi_{\alpha}$
. The parameters $\xi_{1,2,3}$ are introduced to get a smooth variations
of $\xi_{\alpha}$ . Similar to the above cited papers, we choose
the parameters of the Gaussian process as follows, $\overline{g}=0$,
$\sigma\left(g\right)=1$ and $\tau_{\xi}=2$ months. Similar to the
$\xi_{\beta}$, the parameter $\xi_{\alpha}$ varies independently
in the northern and southern hemispheres.

\subsection{Basic parameters and boundary conditions}

The model considers the effects of the large-scale magnetic
field and flow on the axisymmetric mean-field heat transport and angular
momentum balance in the solar convection zone. This part of the model
was described earlier by \citet{Pipin2017} and \citet{Pipin2020}.
Appendix B gives the basic equations governing the angular momentum
balance and the meridional circulation. We use the MESA model \citep{Paxton2011,Paxton2013}
to calculate the reference profiles of mean thermodynamic parameters,
such as entropy, density, temperature and the convective turnover
time, $\tau_{c}$. It assumes that $\tau_{c}$ does not depend
on evolution of the magnetic field and global flows. To define the convective
RMS velocity $u_{c}$ we use the mixing-length approximation,
\begin{equation}
\mathrm{u_{c}=\frac{\ell_{c}}{2}\sqrt{-\frac{g}{2c_{p}}\frac{\partial\overline{s}}{\partial r}},}\label{eq:uc}
\end{equation}
where $\ell_{c}=\alpha_{MLT}H_{p}$ is the mixing length, $\alpha_{MLT}=1.9$
is the mixing length parameter, and $H_{p}$ is the pressure height
scale. The Eq.~(\ref{eq:uc}) defines the profiles of the eddy heat
conductivity, $\chi_{T}$, eddy viscosity, $\nu_{T}$, and eddy diffusivity,
$\eta_{T}$, as follows, 
\begin{eqnarray}
\chi_{T} & = & \frac{\ell^{2}}{6}\sqrt{-\frac{g}{2c_{p}}\frac{\partial\overline{s}}{\partial r}},\label{eq:ch}\\
\nu_{T} & = & \mathrm{Pr}_{T}\chi_{T},\label{eq:nu}\\
\eta_{T} & = & \mathrm{Pm_{T}\nu_{T}}.\label{eq:et}
\end{eqnarray}
The mean entropy, $\overline{s}$, is determined from the mean-field
heat transport equation. The model shows an agreement of the angular
velocity profile with helioseismology results for $\mathrm{Pr}_{T}=3/4$.
The dynamo cycle period of $22$ years is reproduced if $\mathrm{Pm}_{T}=10$
and $C_{\alpha}=0.04$2. The level $C_{\alpha}$ is slightly above
the critical threshold.

We divide the integration domain into two parts. The overshoot region
includes the part of the radiative zone. We put the bottom of the integration
domain at $r_{i}=0.68$R. The convection zone extends from
$r_{b}=0.728R$ to $r_{t}=0.99R$. The solution of the heat transport
gives the mean entropy distribution and determines the turbulent parameters
in the convection zone. In the overshoot region, the intensity of
the turbulent mixing decays exponentially from the bottom of the convection
zone. The bottom boundary rotates as a solid body at the rate
$\Omega_{0}=430$Nhz. At the bottom we put the magnetic field induction
vector to zero. At the top boundary, we use the black-body radiation
heat flux and the stress-free condition for the hydrodynamic part
of the problem.

For the dynamo problem, following ideas of \citet{Moss1992} and \citet{Pipin2011a},
we use the top boundary condition in the form that allows penetration
of the toroidal magnetic field to the surface: 
\begin{eqnarray}
\delta\frac{\eta_{T}}{r_{\mathrm{t}}}B\left(1+\left(\frac{\left|B\right|}{B_{\mathrm{esq}}}\right)\right)+\left(1-\delta\right)\mathcal{E}_{\theta} & = & 0,\label{eq:tor-vac}
\end{eqnarray}
where $r_{\mathrm{t}}=0.99R$. For the set of parameters $\delta=0.999$
and $B_{\mathrm{esq}}=5$G we get the surface toroidal field of magnitude
around 1.5 G. This is in agreement with the results of the solar observations
of \citet{Vidotto2018}. The magnetic field potential outside the
domain is 
\begin{equation}
A^{(vac)}\left(r,\mu\right)=\sum a_{n}\left(\frac{r_{\mathrm{t}}}{r}\right)^{n}\sqrt{1-\mu^{2}}P_{n}^{1}\left(\mu\right),\label{eq:vac-dec}
\end{equation}
where $\mu=\cos\theta$. The nonaxisymmetric part of the dynamo model
was solved using the spherical harmonics. For the numerical solution,
we employ the \textsc{fortran} version of the \textsc{shtns} library
of \citet{shtns}.

\begin{figure}
\centering \includegraphics[width=0.95\columnwidth]{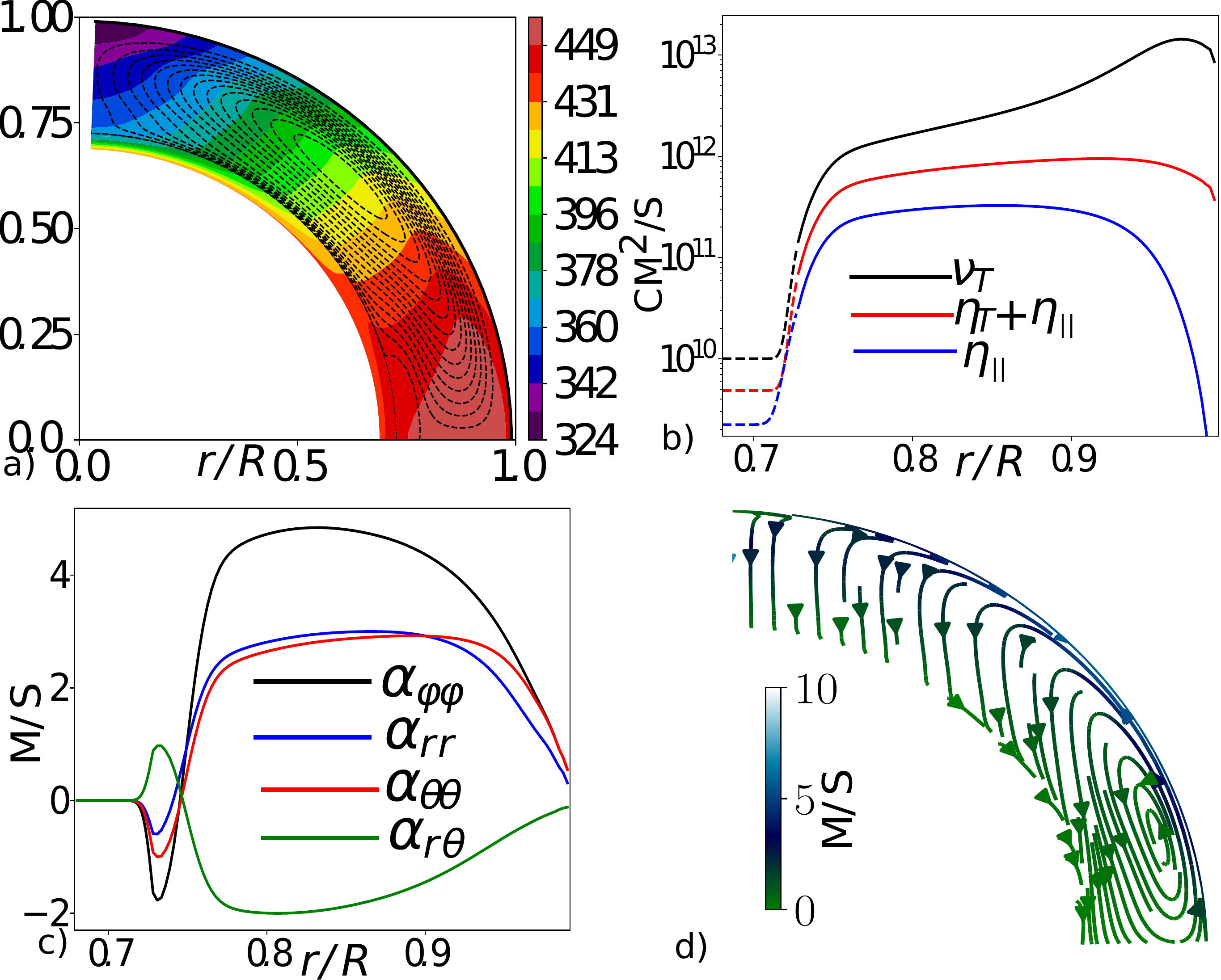} \caption{\label{fig1} a) The meridional circulation (streamlines) and the
angular velocity distributions; the magnitude of circulation velocity
is of 13 m/s on the surface at the latitude of 45$^{\circ}$; b) radial
dependencies of the total, $\eta_{T}+\eta_{||}$, and the rotationally
induced part, $\eta_{||}$, of the eddy magnetic diffusivity and the
eddy viscosity profile, $\nu_{T}$; c) the $\alpha$-effect tensor
distributions at the latitude of 45$^{\circ}$; and d) the streamlines
of the toroidal magnetic field effective drift velocity because of
the meridional circulation and the turbulent pumping effect. Here,
and below we use \textsc{numpy/scipy} \citep{harris2020array,2020SciPyNMeth}
together with \textsc{matplotlib \citep{Hunter2007} }and\textsc{
pyvista \citep{sullivan2019pyvista} }for post-processing and visualization
purposes.}
\end{figure}

The Figure\ref{fig1} illustrates distributions of the angular velocity,
meridional circulation, the $\alpha$ - effect and the eddy diffusivity
in the nonmagnetic case model. The amplitude of the meridional circulation
on the surface is about $13$~m/s. The angular velocity distribution
is in agreement with the helioseismology data.

\begin{table}
\caption{\label{tab}Basic parameters of the reference axisymmetric dynamo
model (run X0, see Table 2) and the BMR's electromotive force}

\begin{tabular}{>{\raggedright}p{0.45\columnwidth}>{\raggedright}p{0.45\columnwidth}}
\hline 
dynamo model parameters  & BMR parameters\tabularnewline
\hline 
$C_{\alpha}=0.042$, $\mathrm{Pr}_{T}=3/4$,$\mathrm{Pm}_{T}=10$,

$\beta^{CZ}\le0.2$, $V_{m}\sim10^{-3}u_{c}$  & emergence time, $\delta t=5$D, growth rate $\tau_{0}=1$D,

$m_{\beta}=100$ (BMR's size $\sim10^{\circ}$),

$V_{\beta}\sim{\displaystyle e^{\delta t/\tau_{0}}V_{m}\sim10\mathrm{m/s}}$,

$\left|\alpha_{\beta}\right|=C_{\alpha\beta}V_{\beta}\sim0-10\mathrm{m/s}$\tabularnewline
\end{tabular}
\end{table}

We list the basic parameters of our dynamo model in the Table\ref{tab}.
The reference axisymmetric model is the same as in the paper of \citet{Pipin2020}.
The axisymmetric dynamo model operates in a weakly nonlinear regime
with $\beta^{CZ}\le0.2$. In this case the maximum of the BMR's
emergency velocity is about 10 m/s. This set a limit for the BMR's
$\alpha_{\beta}$ effect in our model. The fast-rotating solar analogs
can have $\beta^{CZ}\sim1$ (see, \citealt{Pipin21c}). We can expect
the larger magnitudes of $V_{\beta}$ and $\alpha_{\beta}$ for this
case.

\begin{table}
\caption{\label{tab2}The parameters of the runs. The first three columns show
the switches controlling the BMR's activity. The fourth column tells
whether the nonaxisymmetric $\alpha$ effect is included. The $\mathrm{F_{T}}$
is the magnitude of the total unsigned magnetic flux of the toroidal
field in the convection zone; $\mathrm{B_{Pol}}$ is the strength
of the polar magnetic field; $P_{\mathrm{cyc}}$ stands for the dynamo
periods.}

\begin{tabular}{>{\raggedright}p{0.5cm}>{\raggedright}p{0.5cm}>{\raggedright}p{0.5cm}>{\raggedright}p{0.5cm}>{\raggedright}p{0.5cm}>{\raggedright}p{1.1cm}>{\raggedright}p{0.75cm}>{\raggedright}p{0.8cm}}
\hline 
 & $\xi_{\beta}$,

Eq. & $\xi_{\alpha}$,

Eq. & $C_{\alpha\beta}$ & $\tilde{\alpha}_{ij}^{(H)}$,

Eq. & $\mathrm{F_{T}}$,{[}MX{]}

$10^{24}$ & $\mathrm{B_{Pol}}$,

{[}G{]} & $\mathrm{P_{cyc}}$,

yr\tabularnewline
\hline 
X0 & 0 & 0 & 0 & 0 & 1.2$\pm$0.2 & 18.6 & 11.2\tabularnewline
X1 & \eqref{xib} & \eqref{xia} & 0.5 & \eqref{alp2d} & 1.3$\pm$

0.15 & 22.2 & 10.5\tabularnewline
X2 & \eqref{xib} & 0 & 0. & \eqref{alp2d} & 1.3$\pm$

0.15 & 22.2 & 11.2\tabularnewline
X3 & \eqref{xib} & 0 & 0.5 & \eqref{alp2d} & 1.3$\pm$

0.15 & 22.2 & 10.5\tabularnewline
X4 & \eqref{xib} & 0 & 0.5 & 0 & 1.2$\pm$

0.1 & 18.8 & 9\tabularnewline
\end{tabular}
\end{table}

\subsection{Diagnostic parameters}

The Table \ref{tab2} summarizes the control and integral dynamo parameters
of the runs. In addition, we consider the following integral parameters
of the magnetic activity: the magnitude of the total magnetic flux
of the toroidal field in the convection zone, 
\begin{equation}
\mathrm{F_{T}}=2\pi\int_{-1}^{1}\int_{r_{b}}^{r_{\mathrm{t}}}\left|\overline{B}_{\phi}\right|\sin\theta r^{2}\mathrm{d}r\mathrm{d}\mu,\label{eq:ft}
\end{equation}
the strength of the polar magnetic field, 
\begin{equation}
\mathrm{B_{pol}}=\frac{1}{2}\left(\left\langle \overline{B}_{r}\right\rangle ^{\theta<20}-\left\langle \overline{B}_{r}\right\rangle ^{\theta>160}\right),\label{eq:bpol}
\end{equation}
where $\theta$ is the polar angle. We do the averaging of the radial
magnetic field over the polar regions higher than 70$^{\circ}$
latitude. Also, we introduce the total flux of the unsigned radial
magnetic field at the surface,

\begin{equation}
\mathrm{F_{R}}=2\pi R^{2}\varoiint\left|\left\langle B_{r}\right\rangle \right|\sin\theta\mathrm{d}\mu\mathrm{d}\phi.\label{eq:fr}
\end{equation}
We characterize the hemispheric asymmetry of magnetic activity by the
parity index \citep{Knobloch1998}. It defines as follows. As the first
step, we calculate the parameters characterizing the energy of the
symmetric and antisymmetric about the equator parts of the surface radial magnetic field:

\begin{eqnarray*}
E^{\left(e\right)} & = & \varoiint\left[\left\langle B_{r}\right\rangle \left(\mu,\phi,t\right)+\left\langle B_{r}\right\rangle \left(-\mu,\phi,t\right)\right]^{2}\mathrm{d}\mu\mathrm{d}\phi,\\
E^{(o)} & = & \varoiint\left[\left\langle B_{r}\right\rangle \left(\mu,\phi,t\right)-\left\langle B_{r}\right\rangle \left(-\mu,\phi,t\right)\right]\mathrm{d}\mu\mathrm{d}\phi.
\end{eqnarray*}
Then, the parity index reads

\begin{equation}
\mathrm{P}=\frac{E^{(e)}-E^{(o)}}{E^{(e)}+E^{(o)}}.\label{eq:parity}
\end{equation}
Below, $\mathrm{\overline{P}}$ and $\mathrm{\tilde{P}}$ stand for the parity parameter of the axisymmetric
and nonaxisymmetric radial magnetic field, respectively.
For the dipole type of the equatorial symmetry of the radial magnetic
field, we get $\mathrm{P}=-1$, and for the quadrupole type of the
equatorial symmetry, we get $\mathrm{P}=1$. If the magnetic activity
concentrates in one hemisphere, we have $P\approx0$.

The emergence of the BMRs results in the excitation of the large-scale
nonaxisymmetric magnetic field. This nonaxisymmetric magnetic field
takes part in the large-scale dynamo as well. Considering the $\alpha$
effect generation of the large-scale poloidal magnetic field via the
azimuthal electromotive force, we can identify several contributions.
The longitudinal average of the $\alpha$ effect terms in $\mathcal{E}_{\phi}$
results to 
\begin{equation}
\mathcal{E}_{\phi}^{(\alpha)}=\overline{\alpha_{\phi\phi}\left\langle B\right\rangle _{\phi}}=\overline{\alpha}_{\phi\phi}B+\overline{\alpha_{\beta}\left\langle B\right\rangle _{\phi}}+\overline{\tilde{\alpha}_{\phi\phi}\tilde{B}_{\phi}},\label{eq:maina}
\end{equation}
where the first term in the RHS is the standard part of the axisymmetric
mean-field dynamos, the second term represents the effect of the BMRs
or the Leighton effects (see, \citealt{Leighton1969}). The third
term results from the longitudinal averaging of the nonaxisymmetric
part of the $\alpha$ effect and the nonaxisymmetric magnetic field.
It provides the coupling between the evolution of the axisymmetric
and nonaxisymmetric magnetic fields \citep{Bigazzi2004,Berdyugina2006}.
The $\tilde{\alpha}_{\phi\phi}$ in the Eq(\ref{eq:maina}) results
from the nonlinear effects of the nonaxisymmetric magnetic fields
on the kinetic and magnetic parts of the $\alpha$ effect tensor in
the Eq(\ref{alp2d}). In the run X3 we deliberately exclude $\tilde{\alpha}_{\phi\phi}$
to see the impact of the BMR's $\alpha$-effect explicitly.

\section{Results}

\begin{figure}
\includegraphics[width=0.95\columnwidth]{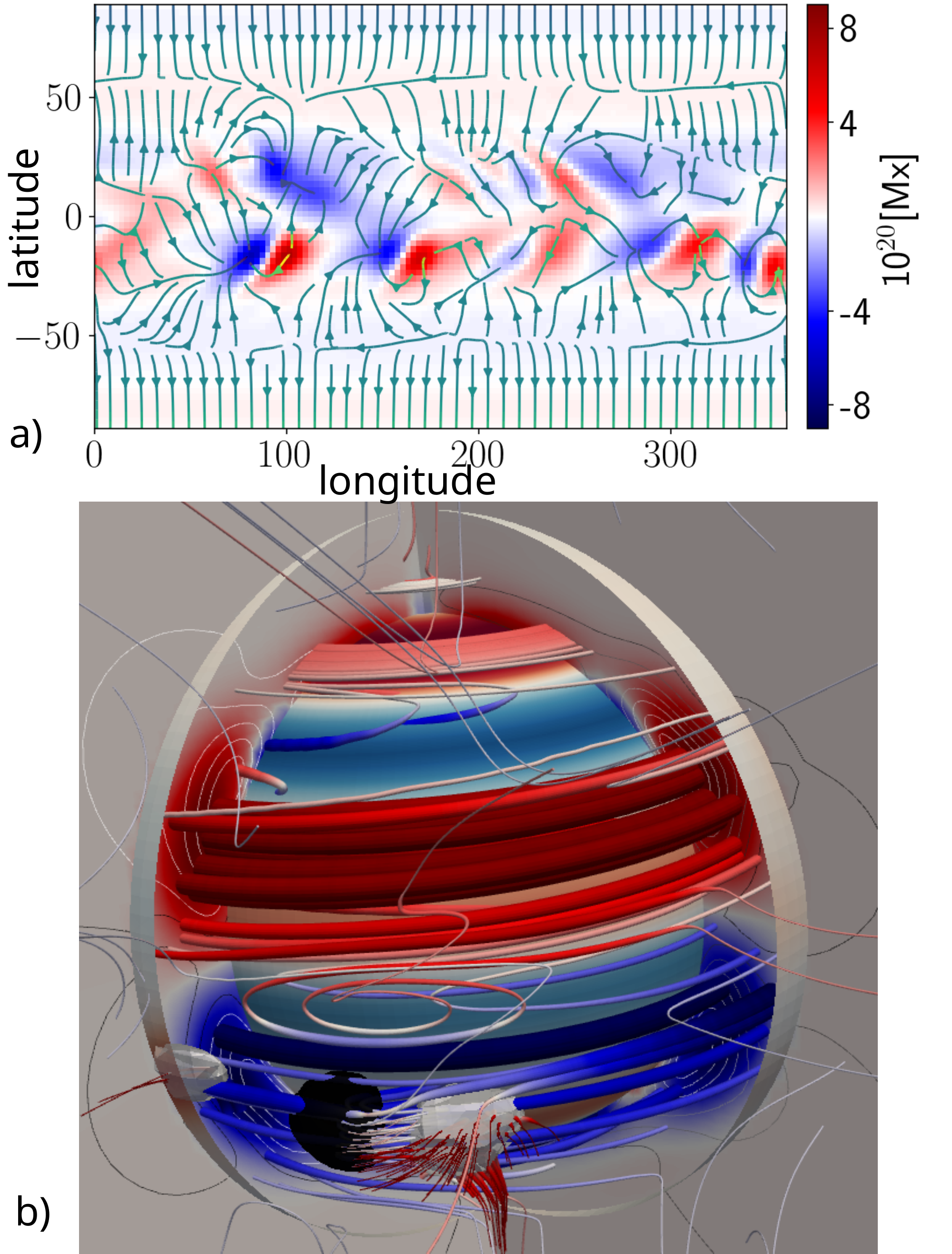}

\caption{\label{fig3}a) The snapshot of the surface radial magnetic flux distribution
and streamlines of the surface magnetic field lines($\left\langle B\right\rangle _{\theta}$,$\left\langle B\right\rangle _{\phi}$);
b) the snapshot of the magnetic field distributions inside the convection
zone; the azimuth of the central meridian is equal to 0; the color
image in the left and right segments and the inner sphere (bottom
of the CZ) shows the toroidal magnetic field strength in the range
of $\pm2$kG; contours in the left and right segments show the streamlines
of the axisymmetric poloidal magnetic field; the magnetic field lines,
which are colored by the gray; red and blue colors reflects the direction
and the magnetic field strength (the white color means $\left\langle B\right\rangle _{\phi}\approx$0);
the BMR is shown by islands which are colored in the black and white
color; the islands confine the volumes of the radial magnetic field
flux above the threshold of $4\cdot10^{20}$Mx per pixel.}
\end{figure}

Figure \ref{fig3} shows snapshots of the magnetic field distribution
during the maximum of the magnetic cycle for model X1. At the low
latitudes of the star, the magnetic field is fairly nonaxisymmetric
in the shallow layer below the surface. The differential rotation
stretches the toroidal field toward the poles from the nonaxisymmetric
remnants of the large unipolar magnetic regions. The animated snapshots
of the magnetic field evolution in run X1 are available online. We
find that our model fairly well reproduces the evolution of the large-scale
magnetic flux on the solar surface. The results are in qualitative
agreement with solar observations (e.g., \citealt{Giovanelli1985,Wang1989b,Virtanen2019,Mordvinov22}).
The distribution of the magnetic field around the BMR is rather shallow
(see, Fig\ref{fig3}b). This is similar to the results of \citet{Miesch2014Ap}.

\subsection{The dynamo effects of BMRs}

\begin{figure}
\includegraphics[width=0.99\columnwidth]{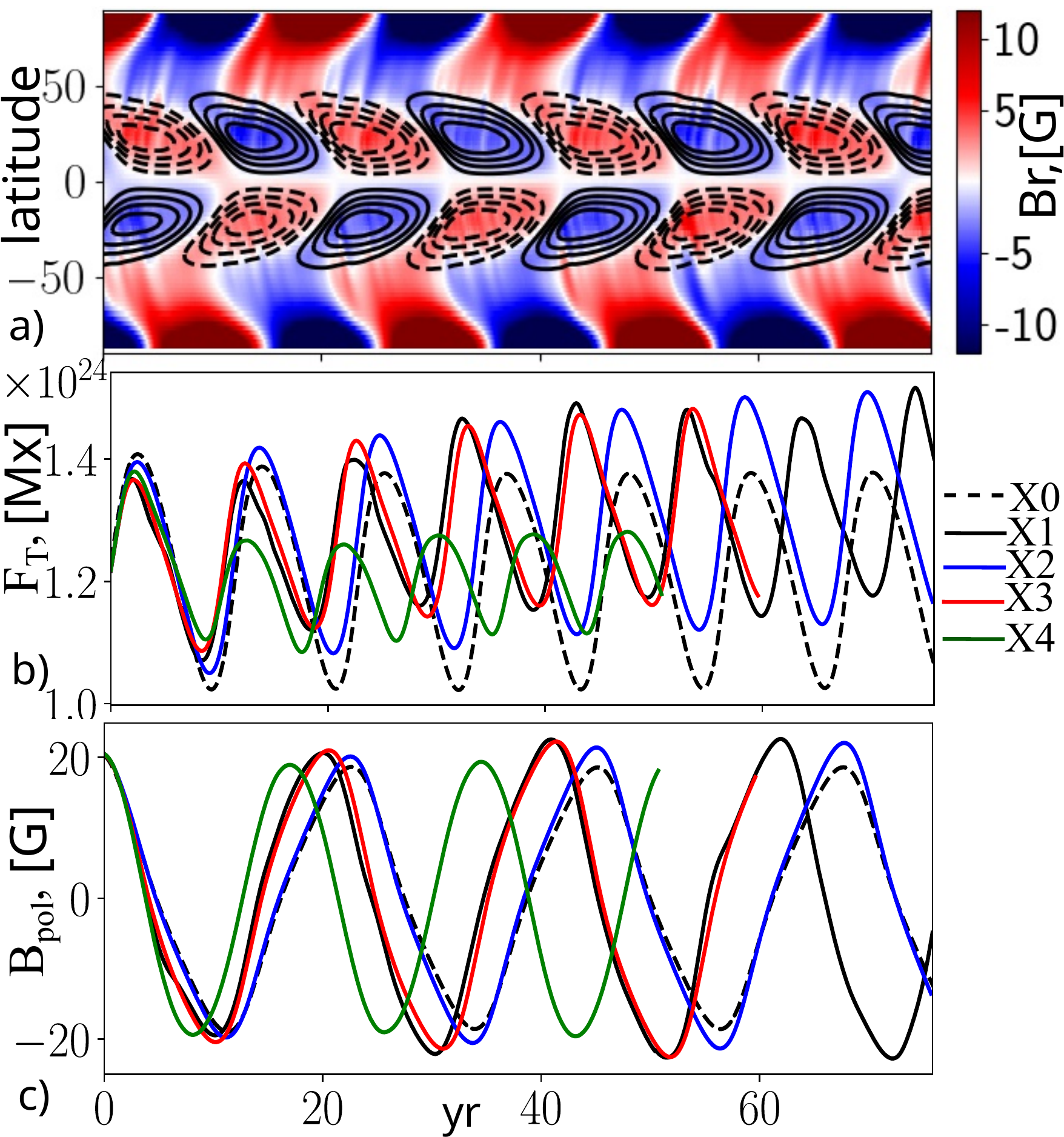}

\caption{\label{fig2}a) The time-latitude diagram of the surface radial magnetic
field (color image) and the toroidal magnetic field at r=0.9R (contours
in range of $\pm1$kG) in the run X1; b) the total flux of the axisymmetric
toroidal magnetic field in the convection zone, $\mathrm{F_{T}}$;
c) evolution of the polar magnetic field, $\mathrm{B_{pol}}$,.}
\end{figure}

Figure \ref{fig2} shows the time-latitude diagram of the axisymmetric
magnetic field evolution in run X1. In the mid-latitudes, the radial
magnetic field shows an intermittent surge-like pattern. It results
from contributions of the weak diffuse field of the large-scale dynamo
and remnants of the BMRs decay. A similar pattern is found in the surface
flux transport models and 3d Babcock-Leighton type models (e.g., \citealt{Mackay2012,Hazra2019n,Kumar2019}).
Also, the intermittent time-latitude pattern of the radial magnetic
field evolution is a typical feature of any 2D dynamo model that includes
the fluctuations of the $\alpha$ effect (see, e.g., \citealt{Yang2020,Pipin2020}).
The other 3D runs listed in Table\ref{tab2}, as well as the axisymmetric
model X0, show the smooth time-latitude diagrams of the radial magnetic
field (see, e.g., \citealt{Pipin2019c}). In our description of the
BMR formation, the BMR's tilt and $\alpha$-effect are readily connected.
Therefore, the fluctuations of the BMR's $\alpha$ effect result in
violations of the Joy law. Our results agree with the conclusions of \citet{Mordvinov22}
who showed that the intermittency of the radial magnetic field evolution
on the surface likely results from violations of Hale's and Joy's
laws in emerging BMRs.

Starting our simulations from the same initial conditions, we compare
the integral dynamo parameters of the runs. Figures \ref{fig2}b and
c show the toroidal flux parameter $\mathrm{F_{T}}$ and the polar
magnetic field, $\mathrm{B_{pol}}$, see the Eqs.(\ref{eq:ft}) and
(\ref{eq:bpol}). All 3D runs, except the run X4 show a higher magnitude
of the $\mathrm{F_{T}}$ and $\mathrm{B_{pol}}$ parameters than the
run X0. The runs X2 and X0 show a longer dynamo period than the runs
X1, X3, and X4. Interesting that the run X4, where we neglect the
mean-field $\alpha$-effect of the nonaxisymmetric magnetic field,
shows the smallest magnitude of the $\mathrm{F_{T}}$ parameter and
the shortest dynamo period. These simulations show the importance
of the mean-field $\alpha$-effect in the dynamo evolution of the
nonaxisymmetric magnetic field. Noteworthy, the Babcock-Leighton solar
dynamo models usually ignore the mean-field $\alpha$ effect acting
on the nonaxisymmetric magnetic fields (terms like $\overline{\tilde{\alpha}_{\phi\phi}\tilde{B}_{\phi}}$
in the Eq(\ref{eq:maina})). We find that the model X2, which has
$\alpha_{\beta}=0$, shows a higher dynamo efficiency than the other
3D runs.

\begin{figure}
\includegraphics[width=0.95\columnwidth]{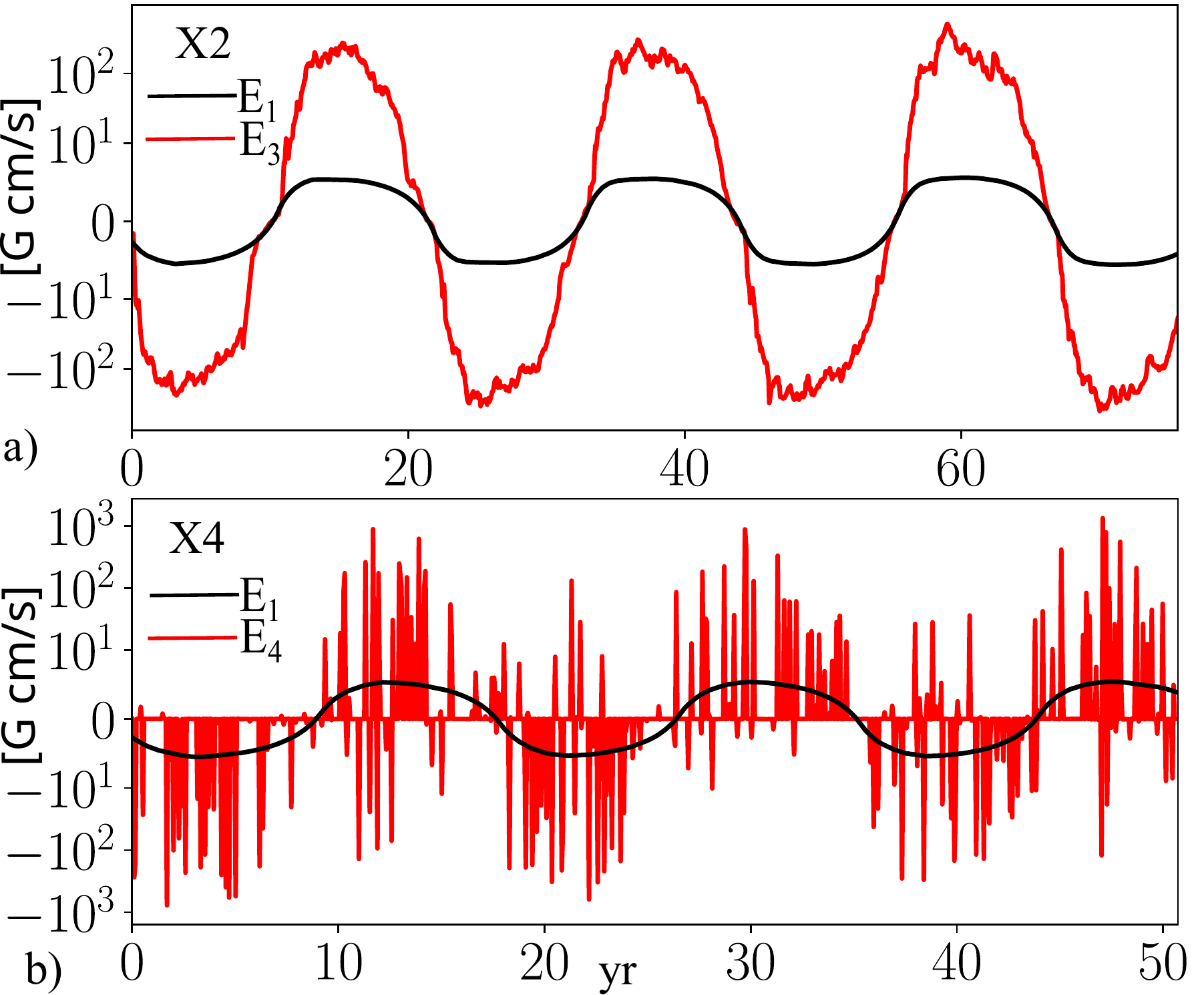}

\caption{\label{fig:harm}The mean electromotive force of the $\alpha$ effect
at the surface: $E_{1}$ - the $\alpha$ effect of the axisymmetric
toroidal field; $E_{3}$ the $\alpha$ effect of the nonaxisymmetric
toroidal field;$E_{4}$- the $\alpha$ effect due to the BMR's tilt.
a) the run X2; b) the same for the run X4.}
\end{figure}

The azimuthal part of the mean electromotive
force, which results in the dynamo generation of the axisymmetric poloidal magnetic field by the $\alpha$
effect, reads,

\begin{align}
\mathcal{E_{\phi}^{\alpha}} & =\left(\overline{\alpha}^{(H)}+\overline{\alpha}^{(M)}\right)B+\overline{\tilde{\alpha}_{\phi\phi}\tilde{B}_{\phi}}+\overline{\alpha_{\beta}\left\langle B\right\rangle _{\phi}},\label{eq:emfal}\\
 & \equiv E_{1}+E_{2}+E_{3}+E_{4}
\end{align}
where the terms $E_{\{n\}}$ denote the corresponding terms in the
Eq(\ref{eq:emfal}), e.g., $E_{2}\equiv\overline{\alpha}^{(M)}B$.
We are interesting to compare the efficiency of the magnetic field
generation by means $E_{3}$ and $E_{4}$. Also, we have to remember
that the mean field generation term $E_{3}$ includes the magnetic
helicity effect (see, the Eq(\ref{alp2d})). Figure \ref{fig:harm}
shows the time series of $E_{1,3,4}$ for the runs X2 and X4 at the
surface. Note that we have $E_{4}=0$ in the run X2 and $E_{3}=0$
in the run X4. The run X2 shows $E_{3}\ge E_{1}$. It is due to $\left|\tilde{B}_{\phi}\right|>\left|\overline{B}_{\phi}\right|$at
the top because of the boundary conditions ($\left|\overline{B}_{\phi}\right|\sim$1G).
The run X4 shows the same. However, the mean variation of the $E_{3}$
in the run X2 shows the greater magnitude than the mean variation
of the $E_{4}$ in the run X4. Because the action of the $\alpha_{\beta}$
is limited by the emerging time of the BMRs. The fluctuations of the
time series $E_{3}$ and $E_{4}$ in the models X2 and X4 are because
of the random emergence of the BMRs at the surface and the magnetic
helicity effect.

Figure \ref{flx}a shows the evolution of the unsigned surface radial
magnetic field flux, $\mathrm{F_{R}}$. In the run X1, the flux correlates
with the activity of the near-surface toroidal magnetic field (cf,
Fig\ref{fig2}a). We find the same in other 3D runs. This agrees with
the solar observations. \citet{Stenflo2013a} found the unsigned surface
radial magnetic field flux to be a good proxy for the sunspot activity.
The runs X1 and X3, which include the $\alpha_{\beta}$ effect, show
the higher magnitude of the $\mathrm{F_{R}}$than the other 3D runs.
These runs show the increase of the baseline of the $\mathrm{F_{R}}$as
well. The run X4 shows the lowest baseline of the $\mathrm{F_{R}}$.
In this run, we neglected the mean-field $\alpha$ effect for the
nonaxisymmetric magnetic field. We conclude that the runs X1-X3 show
the dynamo instability of the nonaxisymmetric magnetic field. This
instability is excited due to the emergence of the BMRs at the top
of the dynamo domain. The run X2 is the interesting case of the dynamo
model run where we neglect the BMR's tilt, putting $\alpha_{\beta}=0$.
Still, in the run X2, the emerging BMRs satisfy, by construction,
the Hale polarity rule. Our results show that, in this case, the mean-field
$\alpha$ effect supports the dynamo generation of the axisymmetric
poloidal magnetic field, and it contributes to the nonaxisymmetric
dynamo, as well.

\subsection{The dynamo parity breaking}

\begin{figure}
\includegraphics[width=0.95\columnwidth]{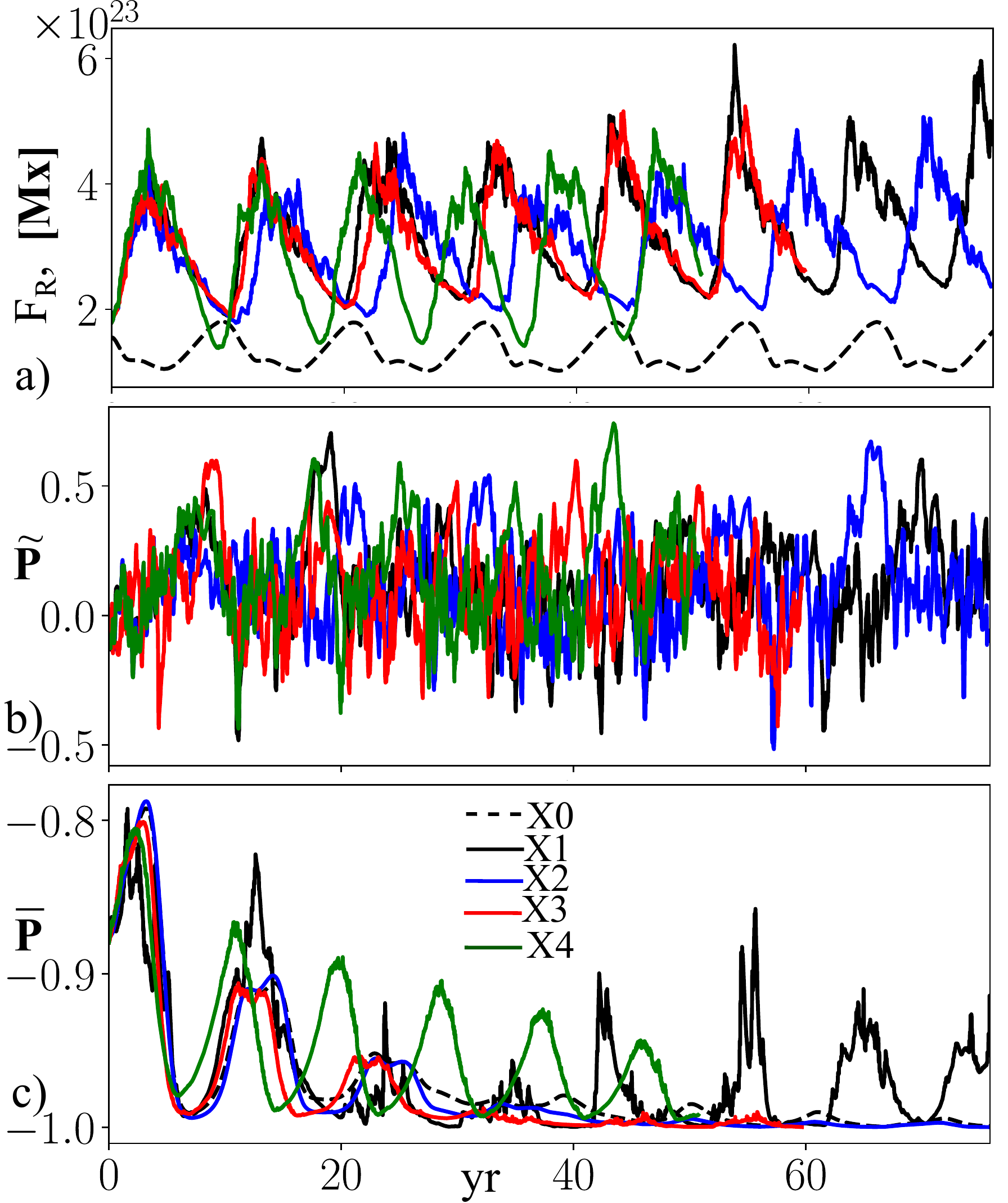}

\caption{\label{flx}a)The total unsigned surface radial magnetic field flux,
$F_{R}$; b) the parity parameter for the nonaxisymmetric radial magnetic
field, $\mathrm{\tilde{P}}$, (see, the Eq\eqref{eq:parity}); c)
the same as b) for the parity parameter of the axisymmetric radial
magnetic field, $\overline{\mathrm{P}}$. The line notation is the
same as in Fig\ref{fig2}.}
\end{figure}

Figs.\ref{flx} b and c show evolution of the parity parameters, $\mathrm{\tilde{P}}$
and $\overline{\mathrm{P}}$. These parameters characterize the equatorial
symmetry of the magnetic activity. The parity of the nonaxisymmetric
magnetic field, $\mathrm{\tilde{P}}$, varies around zero. In the
3D runs the zero magnitude of $\mathrm{\tilde{P}}$ correlates approximately
with epochs of the magnetic activity maximum, which corresponds to
the maxims of the $\mathrm{F_{R}}$ parameter. The maximum magnitude
of the $\tilde{\mathrm{P}}$ (of either positive or negative signs)
correlates approximately with minims of of the $\mathrm{F_{R}}$.
The parameter $\mathrm{\overline{P}}$ shows variations around $-1$,
where $\mathrm{\overline{P}}=-1$ corresponds to the minimum of the
magnetic cycle. In whole, the 3D runs show the highest hemispheric
asymmetry of the magnetic activity for epochs of the magnetic cycle
maximum. The run X1 shows the highest hemispheric asymmetry of magnetic
activity among our runs. 
\begin{figure}
\includegraphics[width=0.95\columnwidth]{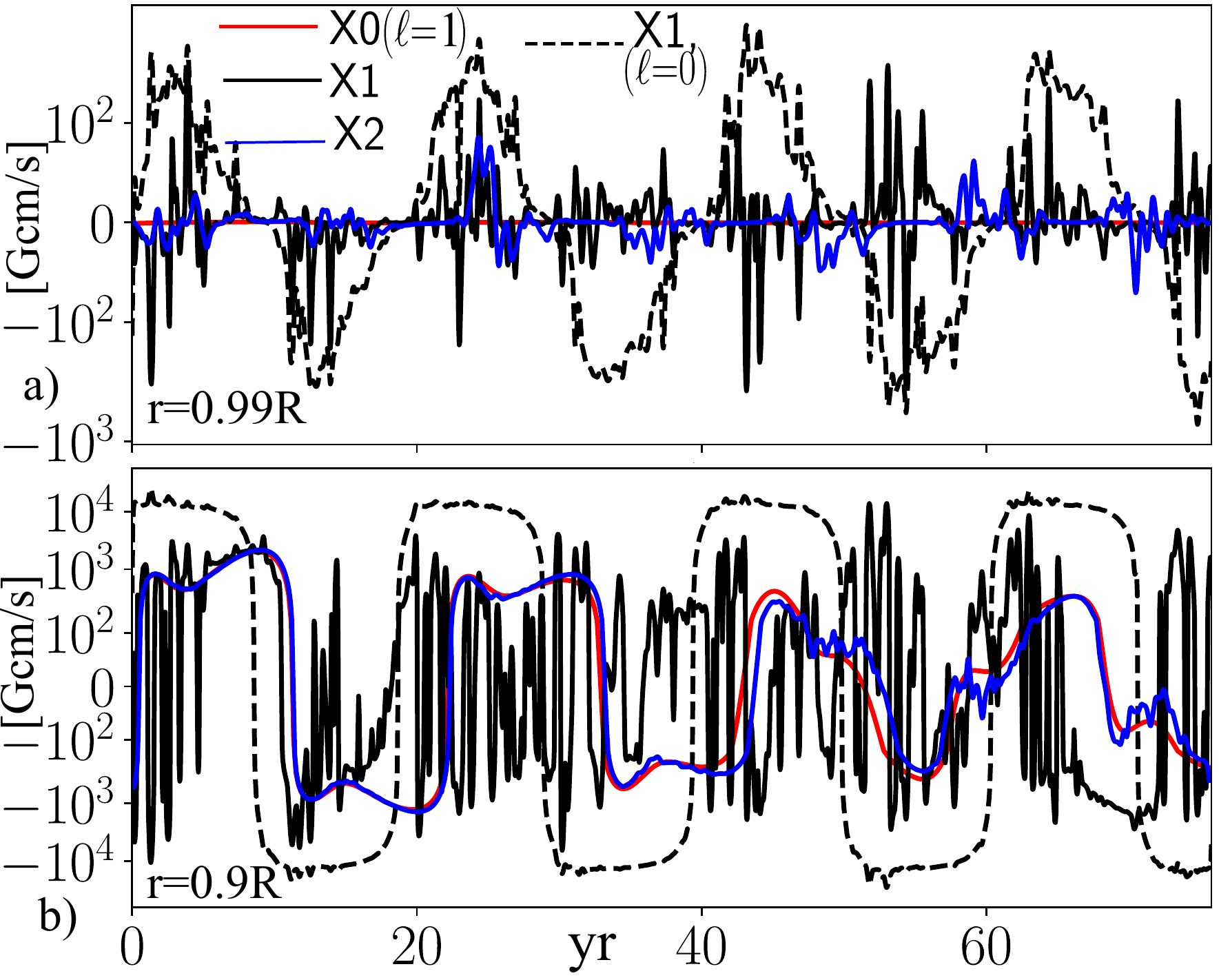}

\caption{\label{fig:harme}a)Evolution of the first three \emph{even} harmonics,
$\ell_{0}$ (black solid line), $\ell_{2}$ (red line), $\ell_{4}$
(blue line) and the odd harmonic $\ell_{1}$ (black dashed line) of
the $\mathcal{E_{\phi}^{\alpha}}$ (see the Eq(\ref{eq:emfal})) at
the near surface level r=0.987R; b) the same as a) for r=0.86R.}
\end{figure}

In our model, the initial magnetic field has a mix of the dipole
and quadrupole parities magnetic fields and the dipole type magnetic
field dominates. We choose the $C_{\alpha}$ parameter about of 10
percent above the dynamo threshold. For this case, the quadrupole
modes of the dynamo solution are below the dynamo instability threshold.
Fig\ref{flx}c shows the slow decay of the quadrupole modes in the
run X0 with time, as the parameter $\overline{\mathrm{P}}$ approaches
to $-1$. The run X1 shows that the fluctuations of the $\alpha_{\beta}$
can support the generation of the quadrupole mode of the large-scale
magnetic field. To shed some light at this phenomenon, we consider
the spherical harmonic decomposition for the mean electromotive force.
The modes $\ell=0$ and $\ell=1$ of the $\mathcal{E}_{\phi}^{(\alpha)}$
(see, the Eq\ref{eq:maina}) generate the dipole and quadrupole modes
of the radial magnetic field, respectively. The Fig\ref{fig:harme}
shows evolution of the modes $\ell=0$ and $\ell=1$ of the $\mathcal{E}_{\phi}^{(\alpha)}$
for the two levels of the convection zone and three dynamo runs, X0,
X1, and X2. The $\ell=0$ mode of the $\mathcal{E}_{\phi}^{(\alpha)}$
shows the stable oscillations of nearly the same magnitude in all
the runs. In the run X1 (with fluctuations of the BMR's tilt), the
mode $\ell=1$ of the $\mathcal{E}_{\phi}^{(\alpha)}$ shows the stable
oscillations both at the surface and inside the convection zone. In
the run X2 there are small fluctuations of the mode $\ell=1$ of the
$\mathcal{E}_{\phi}^{(\alpha)}$ at the surface. They are because
of the magnetic helicity effect and the random emergence of the BMRs.
Inside the convection zone, the run X2 shows a slow decay of the $\ell=1$
mode. Its evolution follows the run X0.

\subsection{Magnetic helicity evolution}

\begin{figure}
\includegraphics[width=0.95\columnwidth]{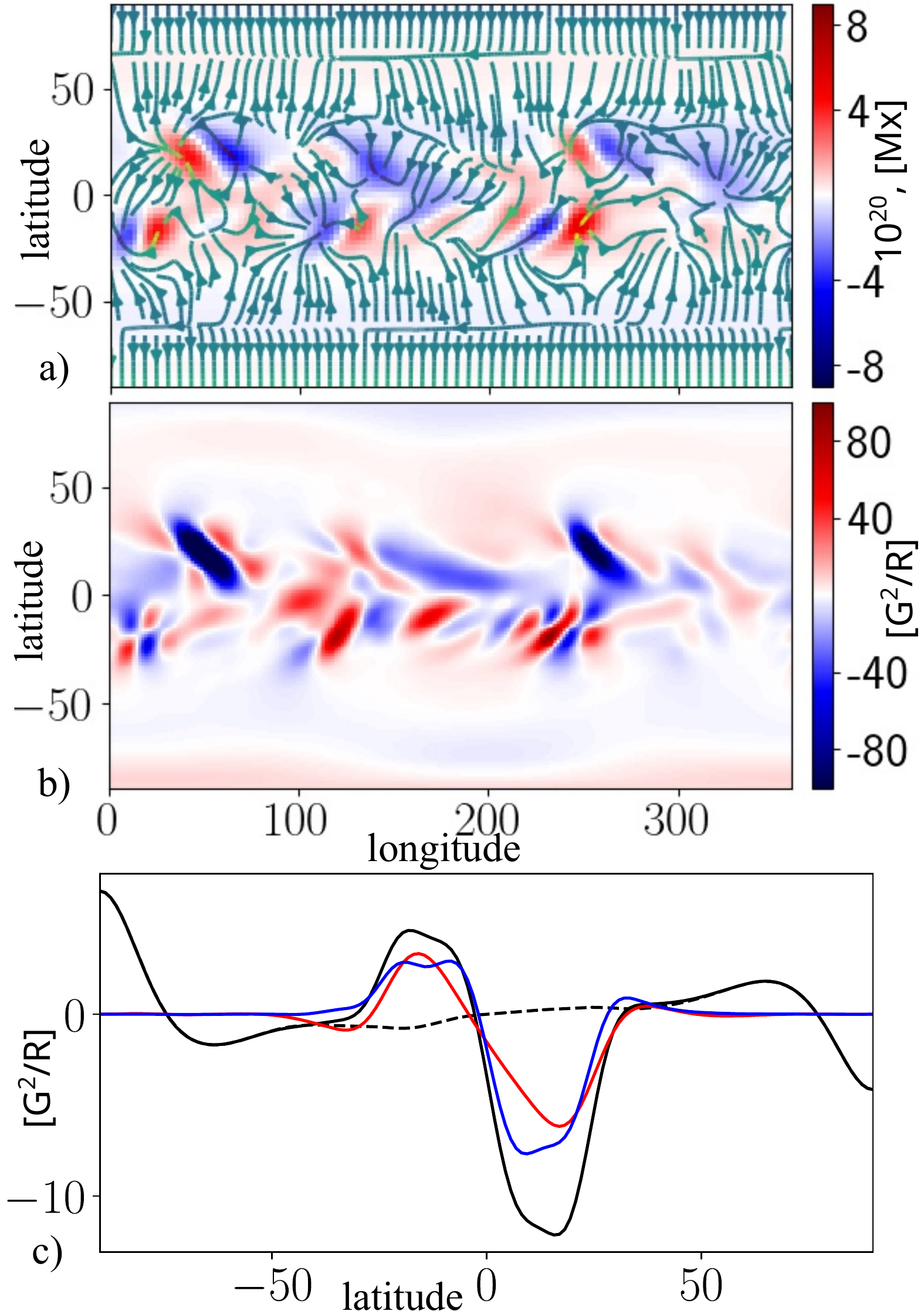}

\caption{\label{fig:held}a)Snapshot of the density flux of the radial magnetic
field (color image) and streamlines of the surface magnetic field
(color measure the strength of $B_{\phi}$ component); b) the snapshot
of the magnetic helicity density; c) the longitudinally averaged magnetic
helicity density: black solid line shows the total magnetic helicity
density $\left\langle \mathbf{A}\right\rangle \cdot\left\langle \mathbf{B}\right\rangle $,
the dashed line shows $\overline{\mathbf{A}}\cdot\overline{\mathbf{B}}$,
the red line - $\tilde{A}_{\varphi}\tilde{B}_{\varphi}$, and the
blue line - $\tilde{A}_{\theta}\tilde{B}_{\theta}$.}
\end{figure}

Figure \ref{fig:held} shows the synoptic map of the surface magnetic
field and magnetic helicity density distribution for the run X1 for
the epoch of the magnetic activity maximum (a month before the snapshot,
which is shown in Fig\ref{fig3}). Similar to the results of \citet{Yeates2020ApJ,Pipin2020b},
we find that the emergence of the BMR in the ambient large-scale field
injects the magnetic helicity at the same place. The effect results
from the magnetic helicity conservation. 

From Fig.\ref{fig:held}b we see that the new active region, which
is at about 20 longitudes in the southern hemisphere, shows the quadrupole
helicity density pattern. The evolution of the emerging BMR involves
the $\alpha$-effect acting on the azimuthal magnetic field surrounding
the BMR and the effect of the differential rotation. Because of the magnetic
helicity conservation, the generated helicity density has the sign
which is opposite to the sign of the $\alpha$-effect. This explain
why the developed BMRs in the southern hemisphere show a preference
for the positive helicity density inside the BMRs. The opposite happens
in the northern hemisphere. 

Because of the total helicity balance, the effect of the injected BMRs
helicity is small. The polar sides of the synoptic
maps show the large-scale field helicity density satisfies the hemispheric
sign rule, i.e., it has a positive sign in the northern hemisphere.
The results of the longitudinal average of the synoptic map Fig\ref{fig:held}b
is shown in Fig\ref{fig:held}c. In each hemisphere, the imbalance
of the total helicity is more than an order of magnitude less than
those extrema in Fig\ref{fig:held}c.

\begin{figure}
\includegraphics[width=0.95\columnwidth]{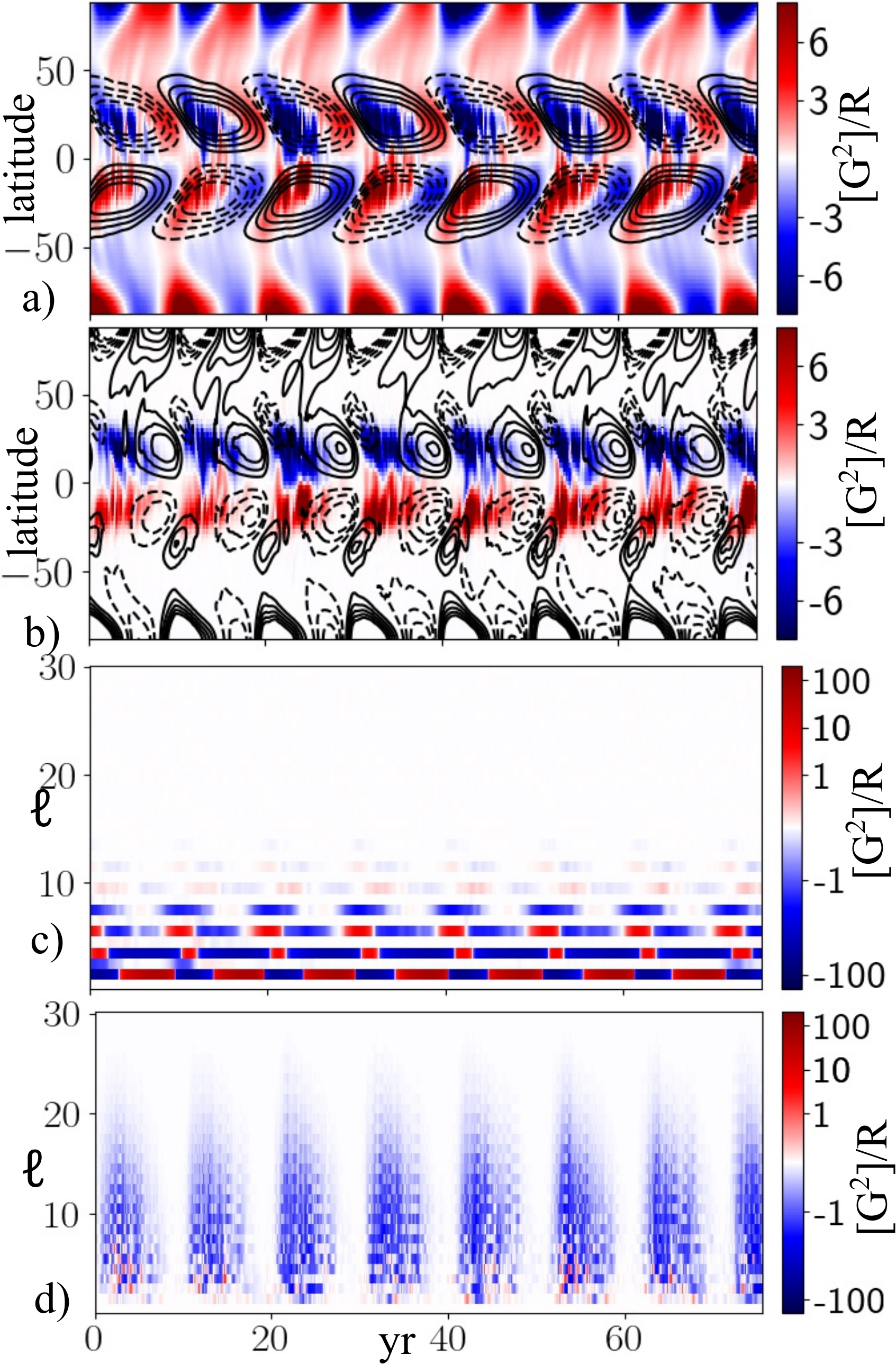}

\caption{\label{fig:hel}a) The time-latitude diagram of the total magnetic
helicity of the large-scale field (color image), contours $\pm1$kG
show the toroidal magnetic field in the subsurface layer r=0.9R; b)
the color image shows the magnetic helicity density of the nonaxisymmetric
magnetic field, and contours show the same (in the same range) for
the axisymmetric magnetic field; c) the helicity proxy spectrum, $H_{\ell}^{+}$
(see, the Eq.\ref{eq:hprx}) evolution of the axisymmetric magnetic
field at the surface; d) the same as c) for the nonaxisymmetric magnetic
field.}
\end{figure}

Fig.\ref{fig:hel} shows results of the model for the time-latitude
evolution of the toroidal magnetic field in the subsurface layer and
the magnetic helicity evolution. The results for the total magnetic
helicity (Fig. \ref{fig:hel}a) are in qualitative agreement with
the solar observation (see, e.g., \citealt{Lund2020} and \citealt{Pevtsov2021}).
The evolution of the small-scale helicity density, $\left\langle \chi\right\rangle $,
which is produced by the dynamo, follows the Eq(\ref{eq:helcon}).
The model shows the surface $\left\langle \chi\right\rangle $ time-latitude
diagram in agreement with the results of our previous papers and we
do not show it here (see, e.g, \citealt{Pipin2013c,Pipin2018b}).
Instead, the rather interesting question is about the helicity density
evolution of the large-scale nonaxisymmetric magnetic field. Fig.\ref{fig:hel}b
show the time-latitude evolution of the $\overline{\tilde{\mathbf{A}}\cdot\tilde{\mathbf{B}}}$
and similar diagram for the axisymmetric magnetic field helicity density,
i.e., $\overline{\mathbf{A}}\cdot\overline{\mathbf{B}}$. The $\overline{\tilde{\mathbf{A}}\cdot\tilde{\mathbf{B}}}$
show the good agreement with the results of \citet{Pipin2019a}, who
measured the helicity density using the synoptic maps of SDO/HMI.
We see that the helicity density $\overline{\tilde{\mathbf{A}}\cdot\tilde{\mathbf{B}}}$
shows the predominantly negative sign in the northern hemisphere and
opposite sign in the southern hemisphere. This agrees with the hemispheric
sign rule (HSR) for the current helicity of the solar active regions
(\citealt{Seehafer1990,Pevtsov1994,Bao2000,Zhang2010}). The magnetic
field, which is involved in the $\overline{\tilde{\mathbf{A}}\cdot\tilde{\mathbf{B}}}$
occupies the intermediate spatial scales. On the surface, the axisymmetric
magnetic field shows variations of the helicity density sign from
the negative in the northern hemisphere during the minimum of the
magnetic cycle to the positive during the maximum cycle. In the southern
hemisphere, the $\overline{\mathbf{A}}\cdot\overline{\mathbf{B}}$
evolves oppositely. A very similar pattern is found in the solar observations
\citep{Pevtsov2021}.

The theoretical expectation suggests the bi-helical magnetic field
in the large-scale dynamo (\citet{Blackman2003,Brandenburg2018,Brandenburg2017}).
The direct computation of the helicity spectrum from distributions
of the vector-potential and magnetic field has two issues. It depends
on the gauge of the vector potential. Another issue is that the different
$\ell$ harmonics of the magnetic helicity density may not show the
same hemispheric sign rule. The two-scale approach of \citet{Roberts1975}
was suggested and developed (see, \citet{Brandenburg2017,Brandenburg2019a})
to overcome the issues. On the first try, we calculate the helicity
proxy spectrum suggested by \citet{Brandenburg2019a}. The proxy is
determined by the spherical harmonics of the superpotentials S and
T (see, \ref{eq:b2}) as follows

\begin{equation}
H_{\ell}^{\pm}=\sum_{m=-\ell}^{m=\ell}2\ell\left(\ell+1\right)S_{\ell\,m}T_{\ell\pm1\,m}^{*},\label{eq:hprx}
\end{equation}
where the shift $\ell\pm1$ was introduced to account for the two-scale
approximation. By definition, the $H_{\ell}^{\pm}$ should have the
positive sign at the low $\ell$ in correspondence of the conventional
hemispheric sign rule, i.e., the predominantly positive helicity for
the large-scale magnetic field in the northern hemisphere. Fig.\ref{fig:hel}c
and d show the evolution of the $H_{\ell}^{+}$ at the surface for
the axisymmetric and nonaxisymmetric parts of the surface magnetic
field. The general sign of $H_{\ell}^{+}$ agrees with the helicity
density evolution shown in Figs\ref{fig:hel}a and b. The $\ell_{1}$
mode shows the positive sign in agreement with the large-scale helicity
density on the synoptic maps of Fig.\ref{fig:held}. 
\begin{figure}
\includegraphics[width=0.8\columnwidth]{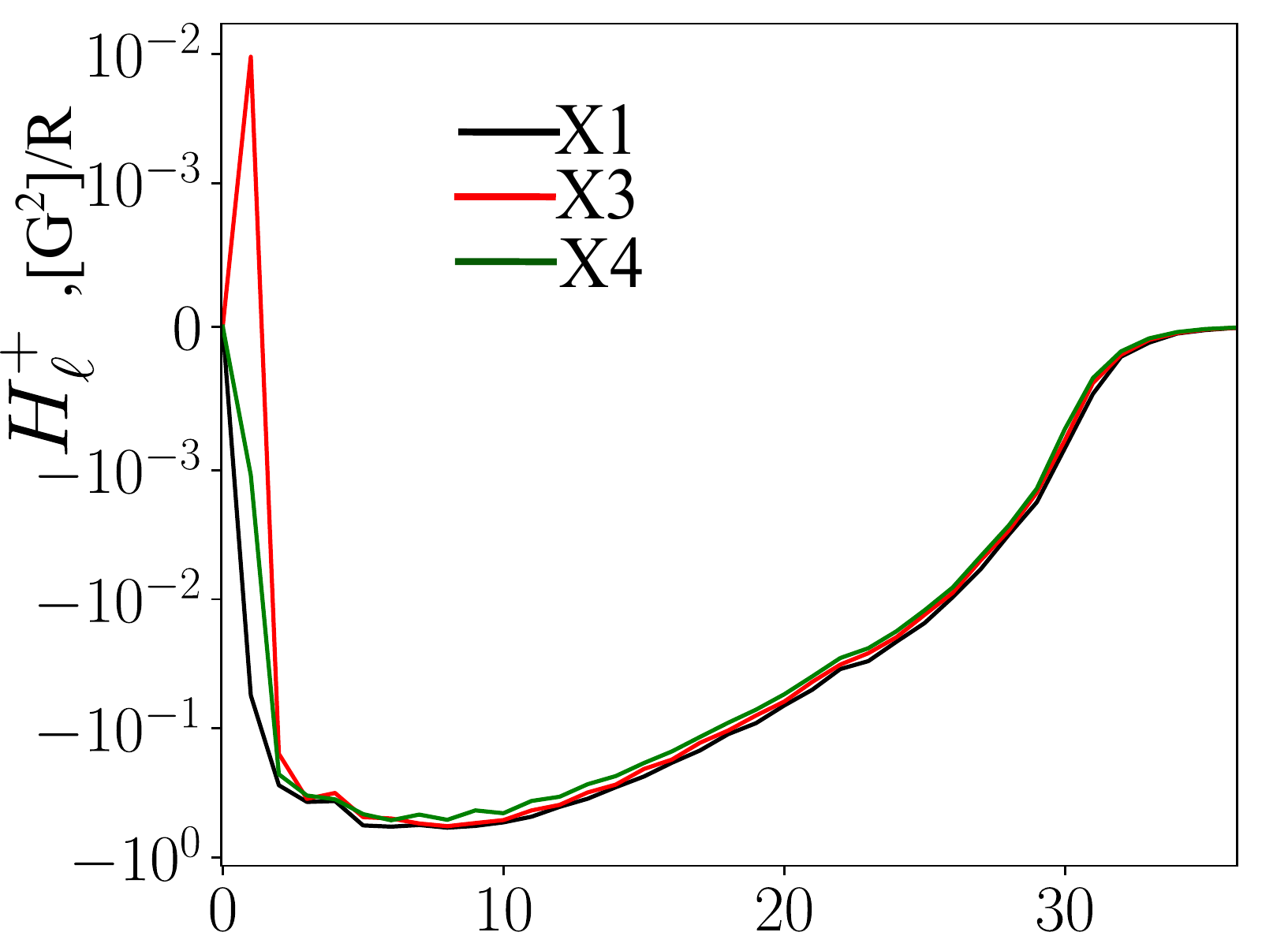}

\caption{\label{fig:helc}The time averaged spectrum $H_{\ell}^{+}$ as computed
from the nonaxisymmetric magnetic field for the runs X1, X3 and X4.}
\end{figure}

Contrary to the theoretical expectations (see, e.g. \citealt{Brandenburg2017}),
we see that $H_{\ell}^{+}$ of the axisymmetric magnetic field shows
the positive sign during the maximum cycle only for the dipole modes.
The octupole modes show the negative sign of $H_{\ell}^{+}$ . Also,
the large-scale nonaxisymmetric field shows a tendency for the positive
helicity sign of $H_{\ell}^{+}$ in the low modes $\ell=1-5$ during
epochs of the magnetic cycle maxima. The sign inversions of $H_{\ell}^{+}$
during those periods of time are present as well. Such a fluctuating
behavior of the bihelical spectrum seems to be typical for the solar
magnetic field \citep{Brandenburg2017,Brandenburg2019a,Prabhu2021}.
In almost all run the bihelical feature of $H_{\ell}^{+}$ disappear
after the averaging of the time series. This is illustrated by Fig\ref{fig:helc}.
Yet, the run X3, which excludes variations of the BMR's tilt, show
$H_{\ell}^{+}>0$ for $\ell=2$ after time averaging. This proves
an important role of the BMR's tilt in the bihelical spectrum of the
large-scale nonaxisymmetric magnetic field of the solar type dynamos. 

\section{Discussion and Conclusions}

The dynamo model, which is presented in the paper, combines the framework
of the mean-field dynamo and phenomenological parts of the Babcock-Leighton
scenarios. The parameters of the mean-field part of our model, such
as the $\alpha$ effect and the eddy diffusivity profiles, are the
same in the axisymmetric and nonaxisymmetric parts of the model. We
calibrate these parameters on the solar observations using the reference
2D dynamo model. The reference dynamo model reproduces the basic properties
of the solar magnetic activity cycle, such as the dynamo cycle period,
the time latitude diagrams of the radial magnetic field evolution.
In addition, the axisymmetric dynamo model consistently reproduces
the solar-like angular velocity and meridional circulation profiles
of the solar convection zone and their solar cycle variations
\citep{Pipin2020,Getling2021}.
The emerging BMRs
affect the solar cycle variations of the angular velocity and the
meridional circulation (\citet{Komm2015,Howe2018,Kosovichev2019,Getling2021}).
We postpone the discussion of these effects to another paper.

The model adopts the Babcock-Leighton scenario using the well-known
effects of the mean electromotive force. The BMR is generated through
the magnetic buoyancy effect. The $\alpha_{\beta}$ effect introduces
the tilt of the BMR. We chose the parameters of the phenomenological
part of the model to fit roughly the parameters of the solar BMRs.
In particular, we show the direct relation of the $\alpha_{\beta}$
effect (see, the Eq\ref{eq:ep}) and the BMR's tilt. Besides the $\alpha_{\beta}$
effect, the model of the BMR's formation includes a lot of free parameters
which control the temporal and spatial characteristics of the BMR's
formation. We did not study the effects of the BMR's formation parameters
on the dynamo evolution in detail.

Compared to the Babcock-Leighton dynamos (e.g., \citealt{Mackay2012,Miesch2014Ap,Kumar2019}),
the effect of the surface magnetic activity on the dynamo in the convection
zone is not strong. The reason is that because the nonaxisymmetric
magnetic field, which is generated from the evolution of the BMRs,
is shallow. For given parameters, the dynamo runs reproduce the total
surface radial magnetic field flux in agreement with observations
by an order of magnitude. The approximate budget of the surface magnetic
flux of the radial magnetic field in our models is of magnitude $\mathrm{F_{R}\sim}5-7\cdot10^{23}$
Mx during the magnetic cycle maximum. Most of this flux is originated
from the emergence and evolution of the BMRs. Comparing the 3D runs
with the axisymmetric dynamo model, we find that this level of the
BMR's activity results in about of 10 percent increase of the total
toroidal magnetic field flux in the convection zone. In the Babcock-Leighton
dynamo models, the tilt of the BMRs is often considered as the
key ingredient for the generation of the large-scale poloidal magnetic
field (see, the above cited paper). Also, the flux-transport dynamo
models ignore the background $\alpha$ effect, which can act on the
nonaxisymmetric magnetic field of the BMRs. We find this mechanism
has a more profound effect on generation of the large-scale poloidal
magnetic field than the BMR's tilt. Therefore, the role of Joy's law
in our dynamo model differs from the pure Babcock-Leighton or flux-transport
dynamo models.

In our model, instead of Joy's law itself, its fluctuations seem to
be rather important. They result in the surge-like pattern of the
poloidal magnetic field evolution on the solar surface. Also, such
fluctuations result in the dynamo parity breaking and the hemispheric
asymmetry of the magnetic activity. \citet{Hazra2019n} discussed
recently in the problem of the long-term parity variations of the
solar magnetic activity. For given dynamo parameters, the axisymmetric
dynamo model shows the decay of the quadrupole modes of the magnetic
field. I find the same for the 3D models that have the fixed BMR's
tilt. Interesting that the fluctuation part of the mean electromotive
force, which generates the quadrupole component of the radial magnetic
field, penetrates deep in the convection zone. Therefore, the surface
effect of the tilt fluctuations of the BMRs results in the global
dynamo excitation of the quadrupole-like magnetic field.

The model allows us to study the effects of the BMR on the surface
magnetic helicity evolution. This is important both for interpretation
of the solar observation and understanding the origin of the helicity
of the solar magnetic field. The emerged BMRs result in surface helicity
density distributions with nearly zero integral imbalance \citep{Yeates2020ApJ}.
Therefore, the magnetic part of the $\alpha$ effect of these BMRs
on the large-scale dynamo is about zero. The results in Fig\ref{fig:hel}
show that on the surface, the effects of the helicity contributions,
which come from the large-scale dynamo, are contaminated by the helicity
injection because of the BMR emergence. The model shows that the hemispheric
sign rule of the large-scale nonaxisymmetric magnetic field helicity
density corresponds to the expected sign rule of the small-scale magnetic
field helicity density, i.e., the $\overline{\tilde{\mathbf{A}}\cdot\tilde{\mathbf{B}}}$
has the negative sign in the northern hemisphere. 

We use the proxy parameter suggested by \citet{Brandenburg2019a}
to investigate the dynamo properties of the magnetic helicity spectrum.
Indeed, the proxy shows the bihelical spectrum (see, Figs\ref{fig:hel}c
and d) with a positive helicity sign for the low $\ell$ modes. The
rest of the spectrum shows the negative helicity sign. Interesting,
that the nonaxisymmetric magnetic field shows a tendency for the bihelical
spectrum during the maximum of the magnetic cycles. This tendency
is rather noisy. The solar observations (see, the above cited papers)
show similar behavior. From a theoretical point of view the bihelical
spectrum of the nonaxisymmetric magnetic field is evidence in favor
of the large-scale nonaxisymmetric dynamo. It is indeed present in
the 3D runs. In our models this dynamo is stochastic by nature. It
is excited by the BMRs activity at the top of the dynamo domain. Our
results shows that the BMR's tilt seems to be important for this stochastic
dynamo. 

In the above discussion, we mentioned that in our model, the dynamo
efficiency of the surface BMRs activity is not strong. Yet, the 2D
dynamo models which neglect this effect can not explain directly the
magnitude of the surface magnetic flux produced by the solar dynamo.
The dynamo model operates in a weakly nonlinear regime with the toroidal
magnetic field strength $\left|\overline{B}\right|<0.2B_{eq}$, where
$B_{eq}$ is the equipartition magnetic field strength. For young
solar analogs which are rotating with a period of 10 days and less,
we expect $\left|\overline{B}\right|\ge B_{eq}$. In this situation,
the BMR's production by the magnetic buoyancy can become much more
efficient than for the modern Sun. Therefore, the surface BMRs activity
can be crucial for the solar-type dynamo in the fast-rotating stars.
The additional dynamo effects, which stem from the BMRs activity,
can decrease the dynamo period (runs X1, X3 and X4). Therefore, this
mechanism can be important in understanding the puzzling behavior
of the dynamo period for the ``quiet'' activity branch of fast-rotating
solar analogs (see, \citealt{Brandenburg2017A}).

Finally, I would like to summarize the major results of the study
as follows. We investigate the effect of the tilted bipolar magnetic
regions (BMR) emergence on the large-scale dynamo distributed in the
bulk of the convection zone. Our results show that bipolar active
regions make a meaningful addition to the dynamo generation of the large-scale
poloidal magnetic field of the Sun. The mean-field $\alpha$ effect,
which acts on the nonaxisymmetric magnetic field of the BMRs, provides
a greater contribution to the dynamo process than the BMR's tilt
does. The fluctuations of the BMR's tilt lead to the parity braking
in the global dynamo. In addition, the helicity density of the nonaxisymmetric
magnetic field of the BMRs shows the hemispheric polarity rule, having
the negative sign in the northern hemisphere of the Sun. Therefore,
the dynamo activity of the surface BMRs seems to control the intermediate
scales of the helicity spectrum of the solar magnetic field at the
photosphere. 

\textbf{Acknowledgements}

This work was carried out within the framework of the international
team project leading by F.A. Pevtsov on Reconstruction of the evolution
of the magnetic field of the Sun and the heliosphere over the past
century with the support of the International Space Science Institute
(ISSI), Bern, Switzerland. Also, the author thanks the financial support
of the Ministry of Science and Higher Education of the Russian Federation
(Subsidy No.075-GZ/C3569/278).

\textbf{Data Availability Statements.} The data underlying this article are
available by request.  
\bibliographystyle{mnras}
\input{buo1.bbl}

\part*{Appendix}

\subsection*{A. The $\alpha$-effect, pumping, and eddy diffusivity.}

The $\alpha$- effect takes into account the kinetic and magnetic
helicities, 
\begin{eqnarray}
\alpha_{ij} & = & 3\eta_{T}C_{\alpha}\psi_{\alpha}(\beta)\alpha_{ij}^{(H)}+\alpha_{ij}^{(M)}\frac{\overline{\chi}\tau_{c}}{4\pi\overline{\rho}\ell^{2}}\label{alp2d-1}
\end{eqnarray}
where $C_{\alpha}$ is a free parameter, the $\alpha_{ij}^{(H)}$
and $\alpha_{ij}^{(M)}$ express the kinetic and magnetic helicity
coefficients, respectively, $\overline{\chi}$- is the small-scale
magnetic helicity, and $\ell$ is the typical length scale of the
turbulence. The helicity coefficients have been derived by \citet{Pipin2008a}.
The $\alpha_{ij}^{(H)}$ reads, 
\begin{eqnarray}
\alpha_{ij}^{(H)} & = & \delta_{ij}\left\{ \left(f_{10}^{(a)}\left(\mathbf{e}\cdot\boldsymbol{\Lambda}^{(\rho)}\right)+f_{11}^{(a)}\left(\mathbf{e}\cdot\boldsymbol{\Lambda}^{(u)}\right)\right)\right\} +\label{eq:alpha}\\
 & + & e_{i}e_{j}\left\{ \left(f_{5}^{(a)}\left(\mathbf{e}\cdot\boldsymbol{\Lambda}^{(\rho)}\right)+f_{4}^{(a)}\left(\mathbf{e}\cdot\boldsymbol{\Lambda}^{(u)}\right)\right)\right\} \nonumber \\
 & + & \left\{ \left(e_{i}\Lambda_{j}^{(\rho)}+e_{j}\Lambda_{i}^{(\rho)}\right)f_{6}^{(a)}+\left(e_{i}\Lambda_{j}^{(u)}+e_{j}\Lambda_{i}^{(u)}\right)f_{8}^{(a)}\right\} ,\nonumber 
\end{eqnarray}
where $\mathbf{e}={\displaystyle \frac{\boldsymbol{\Omega}}{\Omega}},$
$\mathbf{\boldsymbol{\Lambda}}^{(\rho)}=\boldsymbol{\nabla}\log\overline{\rho}$
, $\mathbf{\boldsymbol{\Lambda}}^{(u)}=\boldsymbol{\nabla}\log\left(\mathrm{u'}\ell\right)$
and the $\alpha_{ij}^{(M)}$ reads: 
\begin{equation}
\alpha_{ij}^{(M)}=2f_{2}^{(a)}\delta_{ij}-2f_{1}^{(a)}e_{i}e_{j},\label{alpM}
\end{equation}
Functions $f_{n}^{(a)}\left(\Omega^{*}\right)$ were defined by \citet{Pipin2008a},
$\Omega^{*}=2\tau_{c}\Omega_{0}$, and $\Omega_{0}/2\pi=432$nHz.
In the simulations, we use the case $\varepsilon=1$ (small-scale
magnetic fields in the background turbulence are in equipartition
with the kinetic velocity fluctuations).

\begin{eqnarray*}
f_{1}^{(a)} & = & \frac{1}{4\Omega^{*\,2}}\left(\left(\Omega^{*\,2}+3\right)\frac{\arctan\Omega^{*}}{\Omega^{*}}-3\right),\\
f_{2}^{(a)} & = & \frac{1}{4\Omega^{*\,2}}\left(\left(\Omega^{*\,2}+1\right)\frac{\arctan\Omega^{*}}{\Omega^{*}}-1\right),\\
f_{4}^{(a)} & = & \frac{1}{6\Omega^{*\,3}}\left(3\left(\Omega^{*4}+6\varepsilon\Omega^{*2}+10\varepsilon-5\right)\frac{\arctan\Omega^{*}}{\Omega^{*}}\right.\\
 &  & \left.-\left((8\varepsilon+5)\Omega^{*2}+30\varepsilon-15\right)\right),\\
f_{5}^{(a)} & = & \frac{1}{3\Omega^{*\,3}}\left(3\left(\Omega^{*4}+3\varepsilon\Omega^{*2}+5(\varepsilon-1)\right)\frac{\arctan\Omega^{*}}{\Omega^{*}}\right.\\
 &  & \left.-\left((4\varepsilon+5)\Omega^{*2}+15(\varepsilon-1)\right)\right),\\
f_{6}^{(a)} & = & -\frac{1}{48\Omega^{*\,3}}\left(3\left(\left(3\varepsilon-11\right)\Omega^{*2}+5\varepsilon-21\right)\frac{\arctan\Omega^{*}}{\Omega^{*}}\right.\\
 &  & \left.-\left(4\left(\varepsilon-3\right)\Omega^{*2}+15\varepsilon-63\right)\right),\\
f_{8}^{(a)} & = & -\frac{1}{12\Omega^{*\,3}}\left(3\left(\left(3\varepsilon+1\right)\Omega^{*2}+4\varepsilon-2\right)\frac{\arctan\Omega^{*}}{\Omega^{*}}\right.\\
 &  & \left.-\left(5\left(\varepsilon+1\right)\Omega^{*2}+12\varepsilon-6\right)\right),\\
f_{10}^{(a)} & = & -\frac{1}{3\Omega^{*\,3}}\left(3\left(\Omega^{*2}+1\right)\left(\Omega^{*2}+\varepsilon-1\right)\frac{\arctan\Omega^{*}}{\Omega^{*}}\right.\\
 &  & \left.-\left(\left(2\varepsilon+1\right)\Omega^{*2}+3\varepsilon-3\right)\right),\\
f_{11}^{(a)} & = & -\frac{1}{6\Omega^{*\,3}}\left(3\left(\Omega^{*2}+1\right)\left(\Omega^{*2}+2\varepsilon-1\right)\frac{\arctan\Omega^{*}}{\Omega^{*}}\right.\\
 &  & \left.-\left(\left(4\varepsilon+1\right)\Omega^{*2}+6\varepsilon-3\right)\right).
\end{eqnarray*}

The magnetic quenching function of the hydrodynamical part of $\alpha$-effect:
\begin{equation}
\psi_{\alpha}=\frac{5}{128\beta^{4}}\left(16\beta^{2}-3-3\left(4\beta^{2}-1\right)\frac{\arctan\left(2\beta\right)}{2\beta}\right).
\end{equation}

In the model we take into account the mean drift of large-scale field
due to the magnetic buoyancy, $\gamma_{ij}^{(buo)}$ and the gradient
of the mean density, $\gamma_{ij}^{(\Lambda\rho)}$: 
\begin{eqnarray}
\gamma_{ij} & = & \gamma_{ij}^{(\Lambda\rho)}+\gamma_{ij}^{(buo)},\nonumber \\
\gamma_{ij}^{(\Lambda\rho)}\!\! & =\!\! & \!\!3\nu_{T}f_{1}^{(a)}\!\!\left\{ \!\left(\!\mathbf{\boldsymbol{\Omega}}\cdot\boldsymbol{\Lambda}^{(\rho)}\!\right)\frac{\Omega_{n}}{\Omega^{2}}\varepsilon_{inj}\!\!-\!\!\frac{\Omega_{j}}{\Omega^{2}}\varepsilon_{inm}\Omega_{n}\Lambda_{m}^{(\rho)}\!\!\!\right\} \label{eq:pump1}\\
\gamma_{ij}^{(buo)} & = & \frac{\alpha_{MLT}u_{c}}{\gamma}\mathcal{H}\left(\beta\right)\hat{r}_{n}\varepsilon_{inj},\nonumber 
\end{eqnarray}
where $\mathrm{\alpha_{MLT}}=1.9$ is the MESA mixing-length theory
parameter, $\gamma$ is the adiabatic law constant, $u_{c}$ is the
convective RMS velocity, and 
\[
\mathcal{H}\left(\beta\right)=\frac{1}{8\beta^{2}}\left(\frac{3}{\beta}\arctan\left(\beta\right)-\frac{\left(5\beta^{2}+3\right)}{\left(1+\beta^{2}\right)^{2}}\right),
\]
(see, \citealt{Kitchatinov1993,Ruediger1995})

We employ the anisotropic diffusion tensor following the formulation
of \citet{Pipin2008a} : 
\begin{eqnarray}
\eta_{ijk} & = & 3\eta_{T}\left\{ \left(2f_{1}^{(a)}-f_{2}^{(d)}\right)\varepsilon_{ijk}+2f_{1}^{(a)}\frac{\Omega_{i}\Omega_{n}}{\Omega^{2}}\varepsilon_{jnk}\right\} \label{eq:diff}
\end{eqnarray}
where 
\[
f_{2}^{(d)}=\frac{1}{4\Omega^{*\,2}}\left(\left(\left(\varepsilon-1\right)\Omega^{*\,2}+3\varepsilon+1\right)\frac{\arctan\left(\Omega^{*}\right)}{\Omega^{*}}-\left(3\varepsilon+1\right)\right),
\]
and $\varepsilon=1$. 

\subsection*{B. The angular momentum balance and the meridional circulation}

Our dynamo model takes into account the effects of the magnetic activity
on the angular momentum balance: 
\begin{eqnarray}
\frac{\partial}{\partial t}\overline{\rho}r^{2}\sin^{2}\theta\Omega\! & = & \!\!-\boldsymbol{\nabla\cdot}\left(r\sin\theta\overline{\rho}\left(\!\hat{\mathbf{T}}_{\phi}\!+r\sin\theta\Omega\mathbf{\overline{U}^{m}}\!\right)\!\right)\label{eq:angm}\\
 & + & \boldsymbol{\nabla\cdot}\left(r\sin\theta\frac{\overline{\left\langle \mathbf{B}\right\rangle \left\langle B\right\rangle _{\phi}}}{4\pi}\right),\nonumber 
\end{eqnarray}
where the overbar means the azimutal averaging, and $\left\langle \mathbf{B}\right\rangle =\overline{\mathbf{B}}+\tilde{\mathbf{B}}$.
Here, $\overline{\mathbf{B}}$ and $\tilde{\mathbf{B}}$ are the axisymmetric
and nonaxisymmetric components of the large-scale magnetic field.
We find that the magnetic tension contribution can be decomposed into
sum: $\overline{\mathbf{B}}\overline{B}_{\phi}+\overline{\tilde{\mathbf{B}}\tilde{B}_{\phi}}$,
where the second part represents the longitudinal average of the magnetic
tensions from the nonaxisymmetric magnetic fields. The meridional
circulation is governed by equation for the azimuthal component of
large-scale vorticity, $\mathrm{\overline{\omega}=\left(\boldsymbol{\nabla}\times\overline{\mathbf{U}}^{m}\right)_{\phi}}$:

\begin{eqnarray}
\mathrm{\frac{\partial\omega}{\partial t}\!\!\!} & \mathrm{\!\!=\!\!\!\!} & \mathrm{r\sin\theta\boldsymbol{\nabla}\cdot\left(\frac{\hat{\boldsymbol{\phi}}\times\boldsymbol{\nabla\cdot}\overline{\rho}\hat{\mathbf{T}}}{r\overline{\rho}\sin\theta}-\frac{\mathbf{\overline{U}}^{m}\overline{\omega}}{r\sin\theta}\right)}\label{eq:vort}\\
 & + & \mathrm{r}\sin\theta\frac{\partial\Omega^{2}}{\partial z}-\mathrm{\frac{g}{c_{p}r}\frac{\partial\overline{s}}{\partial\theta}}\nonumber \\
 & + & \frac{1}{4\pi\overline{\rho}}\overline{\left(\mathbf{\left\langle B\right\rangle }\boldsymbol{\cdot\nabla}\right)\left(\boldsymbol{\nabla}\times\left\langle \mathbf{B}\right\rangle \right)_{\phi}}\nonumber \\
 &  & -\frac{1}{4\pi\overline{\rho}}\overline{\left(\left(\boldsymbol{\nabla}\times\left\langle \mathbf{B}\right\rangle \right)\boldsymbol{\cdot\nabla}\right)\left\langle \mathbf{B}\right\rangle }_{\phi},\nonumber 
\end{eqnarray}
where $\hat{\mathbf{T}}$ is the turbulent stress tensor. Also, $\overline{\rho}$
is the mean density, $\mathrm{\overline{s}}$ is the mean entropy;
$\mathrm{\partial/\partial z=\cos\theta\partial/\partial r-\sin\theta/r\cdot\partial/\partial\theta}$
is the gradient along the axis of rotation. The second line accounts
for the source terms of the meridional circulation, which are due
to the centrifugal and baroclinic forces. We neglect the effects of
the rotational oblateness of the density and pressure profiles. More
details about this part of the model as well as the model of the mean-field
heat transport can be found in \citet{Pipin2019c}. We plan to discuss
the effects of the magnetic tensions, which results from the BMRs
evolution, in a separate paper. 
\end{document}